\documentclass[10pt]{article}
\usepackage{amsfonts,amssymb,amsmath,empheq}
\usepackage{amsthm}
\usepackage{booktabs}
\usepackage[toc,page]{appendix}
\usepackage{graphics}
\usepackage{graphicx,color}
\usepackage{latexsym}
\usepackage{subfig}
\usepackage{multirow}
\usepackage{float}
\usepackage{cases}
\usepackage{bm}
\usepackage[ruled]{algorithm2e}
\usepackage{xcolor}
\usepackage[normalem]{ulem}
\usepackage{indentfirst}
\usepackage{epstopdf}
\usepackage{CJK}
\usepackage[colorlinks,
            linkcolor=red,
            anchorcolor=blue,
            citecolor=green
            ]{hyperref}
\theoremstyle{plain}
\numberwithin{equation}{section}

\newcommand{\beqa}{\begin{eqnarray}}
\newcommand{\eeqa}{\end{eqnarray}}

\newcommand{\bomega}{\mbox{\boldmath $\omega$}}

\newcommand{\beq}{\begin{equation}}
\newcommand{\eeq}{\end{equation}}
\def\R{{\mathbb{R}}}

\def\bA{\mathbf{A}}
\def\bB{\mathbf{B}}
\def\bG{\mathbf{G}}
\def\bW{\mathbf{W}}

\def\bL{\mathbf{L}}

\def\f{\mathbf{f}}

\def\bc{\mathbf{c}}
\def\bg{\mathbf{g}}
\def\bs{\mathbf{s}}

\def\by{\mathbf{y}}

\def\bx{\mathbf{x}}
\def\bM{\mathbf{M}}

\def\bL{\mathbf{L}}
\def\bo{\mathbf{0}}
\def\bI{\mathbf{I}}

\def\R{\mathbb{R}}
\def\fhat{\hat{\mathbf{f}}}
\newcommand{\norm}[1]{\left\lVert#1\right\rVert}
\def\argmin{\mathop{\rm argmin}}

\newcommand{\yob}{\mathbf{y}_{\text{ob}}}
\newcommand{\lp}{\left(}
\newcommand{\rp}{\right)}
\newcommand{\lb}{\left[}
\newcommand{\rb}{\right]}
\newcommand{\lc}{\left\{}
\newcommand{\rc}{\right\}}

\newenvironment{sciabstract}{%
\begin{quote} \bf}
{\end{quote}}

\makeatletter
\newcommand*{\addFileDependency}[1]{
  \typeout{(#1)}
  \@addtofilelist{#1}
  \IfFileExists{#1}{}{\typeout{No file #1.}}
}
\makeatother

\usepackage{authblk}
\title{Span of regularization for solution of inverse problems, with application to magnetic resonance relaxometry} 
\author[1]{Chuan Bi}
\author[2]{Yvonne M. Ou}
\author[1]{Mustapha Bouhrara}
\author[1]{Richard G. Spencer}
\affil[1]{National Institute on Aging, Baltimore, MD 21224, U.S.A}
\affil[2]{Department of Mathematical Sciences, University of Delaware, Newark, DE 19716, U.S.A}
\affil[ ]{\textit {chuan.bi@nih.gov, mou@udel.edu, mustapha.bouhrara2@nih.gov, spencerri@mail.nih.gov}}
\date{}
\begin{document}
\maketitle 

\begin{sciabstract} 
    We present a fundamentally new regularization method for the solution of the Fredholm integral equation of the first kind, in which we incorporate solutions corresponding to a range of Tikhonov regularizers into the end result. This method identifies solutions within a much larger function space, spanned by this set of regularized solutions, than is available to conventional regularization methods. Each of these solutions is regularized to a different extent. In effect, we combine the stability of solutions with greater degrees of regularization with the resolution of those that are less regularized. In contrast, current methods involve selection of a single, or in some cases several, regularization parameters that define an optimal degree of regularization. Because the identified solution is within the span of a set of differently-regularized solutions, we call this method \textit{span of regularization}, or SpanReg. We demonstrate the performance of SpanReg through a non-negative least squares analysis employing a Gaussian basis, and demonstrate the improved recovery of bimodal Gaussian distribution functions as compared to conventional methods. We also demonstrate that this method exhibits decreased dependence of the end result on the optimality of regularization parameter selection.  We further illustrate the method with an application to myelin water fraction mapping in the human brain from experimental magnetic resonance imaging relaxometry data. We expect SpanReg to be widely applicable as an effective new method for regularization of inverse problems.

\end{sciabstract}
\newpage
\section*{Significance Statement}
Estimating model parameters from data is notoriously difficult, with results that can exhibit extreme sensitivity to noise. There is a large literature on limiting this sensitivity through methods known as regularization; these require an optimal amount of regularization to be defined so that the derived solution is not over-smoothed. All non-optimal solutions are discarded.  However, there is substantial information content in these discarded solutions.  Accordingly, we present a fundamentally new method, SpanReg, that incorporates solutions for a range of regularizations into the end result and shows substantial improvements over existing methods. We apply this to magnetic resonance imaging myelin mapping in the human brain. We expect SpanReg to be widely applicable as an effective method for regularization.

\section{Introduction}
\subsection{\label{discrete_setup}Regularization of Inverse Problems Arising from the Fredholm Equation of the First Kind}
A large class of inverse problems arise from a discretized Fredholm integral equation of the first kind. With the introduction of upper and lower bounds based on prior knowledge, this may be written as 
\begin{equation}
\label{continuous_fwd_prob}
    y(t) = \int_{\tau_{L}}^{\tau_{U}} A(t,\tau) f(\tau) d\tau
\end{equation}
where y(t) is the measurement or observation, $A(t,\tau)$ is called the \textit{kernel} and $f(\tau)$ is the distribution function (DF) to be determined. A discretization of Eq. \eqref{continuous_fwd_prob} leads to
\begin{equation}
    \label{discrete_fredhlom}
    y(t_i) = \sum_{j = 1}^n A(t_i, \tau_j) f(\tau_j) \Delta \tau,\quad i = 1,2,\cdots, m
\end{equation}
and
\begin{equation}
    \label{discrete_fredholm_matrix_form}
    \by = \bA \f
\end{equation}
where $\by \in \R^m$, $\bA \in \R^{m\times n}$ with $A_{i,j} := A(t_i, \tau_j)\Delta \tau$, and $\f \in \R^n$. Here we assume a uniform discretization $\triangle\tau:=\frac{\tau_{L}-\tau_U}{n}$ and $\tau_i=\tau_L+(i-1)\triangle\tau$ for convenience. Depending on the application, soft or hard constraints on $f(\tau)$ may include, for example, degree of smoothness, values for upper and lower bounds, and amount of total variation.

One important application of this formulation is to magnetic resonance relaxometry (MRR), which estimates the distribution of relaxation times $T_2$ within a sample. We emphasize that this serves as a paradigm for a much larger class of physical phenomena leading to essentially identical mathematical considerations. Physically, $T_2$, also called the transverse relaxation time, represents the time constant for decay of transverse magnetization. The range of possible $T_2$ values depends on the sample or tissue under study, but typical values may range from $\sim$10 ms to 2000 ms in biomedical studies, with smaller values corresponding to more solid or rigid tissues. Shorter values require specialized techniques to detect, depending on available hardware and experimental protocols.  The distribution function $f(T_2)$ of such values can be of great importance in characterizing the physical properties and the material composition of a sample. In MRR, the kernel function is of the special form $A(t,T_2) = \exp(-\frac{t}{T_2})$ and the Fredholm integral equation in \eqref{continuous_fwd_prob} is a Laplace transform. Explicitly, the signal model is 
\begin{equation}
    \label{laplace_transform_continuous}
    y(t) = \int e^{-\frac{t}{T_2}} f(T_2)d T_2.
\end{equation}
This defines the integrated signal from an ensemble of decaying exponentials, with the contribution for a given value of the decay constant $T_2$ determined by $f(T_2)$. Thus, as the time $t$ increases, the overall signal decays into the noise with a mixture of time constants, the distribution of which is to be determined.  Practically, the observed data $\yob \in \R^m$ is obtained at measurement times $\{t_i\}_{i=1}^m$, and the goal is to determine the $T_2$ distribution $\f \in \R^n$ as defined by the discretized problem:  
\begin{equation}
\by_{\text{ob}} = \bA \f + \bomega, \quad \f \geq \bo
\label{fwd_eqn}
\end{equation}
where $\bA\in \R^{m\times n}$ is the kernel matrix with entries $A_{i,j} = e^{-\frac{t_i}{T_{2,j}}} \Delta T_2$ and $\bomega$ is additive random noise. We define the SNR of the signal by
\begin{equation}
    \text{SNR} = \frac{\max \left| \by_{\text{ob}} \right|}{\text{RMS}(\bomega)}
    \label{SNR_def}
\end{equation}
where RMS$(\bomega)$ is the root mean square noise amplitude.  The positivity constraint $\f \geq \bo$ arises from the physical requirement that $\f$ represents the volume fraction of materials in the sample. The matrix $\bA$ inherits the smoothing property of the integral operator in \eqref{continuous_fwd_prob}, and can exhibit a condition number that is so large that any direct inversion of the noisy data $\by_{ob}$ without regularization, such as through non-negative least squares (NNLS) analysis \cite{Spencer_Bi_2020}, can be extremely unstable.

From the perspective of inverse problems, determination of the discretized $T_2$ distribution $\f$ can be seen as a version of the inverse Laplace transform (ILT), a classic ill-posed problem. An important method for addressing the numerical instability inherent in this process is through Tikhonov regularization.  In the context of NNLS, where $\f$ is everywhere non-negative, the native problem
\begin{equation}\label{NNLS}
\f_{LS} = \argmin_{\f \geq \bo}\norm{\bA \f - \by_{\text{ob}}}_2^2
\end{equation}
is replaced by the closely related problem
\begin{equation}
\f_{\lambda} = \argmin_{\fhat \geq \bo} \left\lbrace \norm{\bA \fhat - \by_{\text{ob}}}_2^2 + \lambda^2\norm{\fhat}_2^2  \right\rbrace.
\label{constrained_tikhonov_regularization}
\end{equation}
The second term serves to penalize large values of the norm of the recovered DF, limiting sensitivity to noise.  The regularization parameter $\lambda$ acts to titrate the relative importance of the two terms in Eq. \eqref{constrained_tikhonov_regularization}. The recovered DF $\fhat$ may be highly dependent on this parameter; selection of an optimal $\lambda$ remains the topic of active research.  This formulation in effect replaces the original ill-posed problem, Eq. \eqref{NNLS}, by Eq. \eqref{constrained_tikhonov_regularization}, a more well-behaved but different problem \cite{hansen2010discrete}. Additionally, prior application-dependent assumptions regarding $\fhat$, such as sparsity or smoothness, can be introduced into the minimization by introduction of suitable alternative or additional regularization terms \cite{sabett2017l1, Spencer_Bi_2020}.  

\subsection{Methods of Parameter Selection}
The problem of selecting $\lambda$ in Tikhonov regularization has been studied for decades, with no universal approach having been identified. Classical methods such as the Morozov discrepancy principle (DP) \cite{morozov2012methods}, the L-curve \cite{hansen_analysis_1992,hansen_rank-deficient_1987,hansen_use_1993,calvetti_tikhonov_2000}, and generalized cross-validation (GCV) \cite{golub1979generalized,wahba1977practical,wahba1990spline} define criteria by which to select $\lambda$. For example, the DP seeks a value of $\lambda$ such that the size of the first term in Eq. \eqref{constrained_tikhonov_regularization}, called the residual, is matched to the noise level in the observed data. The L-curve method identifies a value of $\lambda$ that defines the corner of the L-shaped curve defined by the log-log plot of the solution norm against the norm of the residual. GCV selects lambda using an expression based on leave-one-out cross-validation.  All of these methods seek to identify one optimal $\lambda$ and its corresponding regularized solution, and discard all other solutions. Variations on this include the elastic net, which incorporates two different $\lambda$'s serving what are essentially distinct roles, but still identifies single optimal values for each \cite{Zou_2005}.  A survey of methods can be found in \cite{bauer_comparing_2011}. 

The central theme of the present work is that discarding all results except the reconstruction corresponding to a single selected $\lambda$ leads to loss of information that could contribute to the accurate recovery of the DF, $\f$. We have observed that the effects of regularization, for example, variations of widths, amplitudes, and shapes of recovered distributions $\f_{\lambda_j}$, depend on the underlying distribution $\f$. This motivates the notion that the solutions corresponding to values of $\lambda$ other than the one selected may contain additional information, so that improved results may be obtained by incorporating these solutions into the final determination of $\f$ Accordingly, we describe a new method, termed \textit{span of regularization, or SpanReg}, for which the recovered distribution function is a linear combination of regularized solutions across a range of $\lambda$'s. This expands the space of functions from which the desired DF is drawn.

\subsection{Gaussian Mixture Representation and Application to Determining $T_2$ Distribution Functions}
We use a Gaussian mixture representation as a dictionary to describe the unknown DF, $\f$, as required for our analysis. One approach would be to write the DF as the linear combination of a finite set of Gaussian functions $g(\mu_i,\sigma_i)$, with $\mu_i$ and $\sigma_i$ representing the unknown mean and standard deviation (SD) of a given element of that set. The discretization of the $g_i$'s along the abscissa follows from the discretization of $\f$ along the abscissa, that is, the choice of the set of abscissa $T_2$ values \cite{Bonny_2020}. With a prior assumption of the number $M$ of Gaussian components required for an adequate description, the determination of $\f$ can be recast as the non-linear least squares problem:
\begin{equation}\nonumber
    \argmin_{\lc\sigma_i, \mu_i\rc}{\norm{\yob - \bA\sum_{i=1}^M \bg(\sigma_i,\mu_i) }_2^2}
\end{equation}
Alternatively, by establishing a dictionary of Gaussian functions of specified $\mu_i$ and $\sigma_i$ \cite{couvreur1996dictionary,han2013gaussian} and incorporating the non-negativity of $\f_{\text{true}}$, we instead have the problem:
\begin{equation}\nonumber
    \argmin_{\bc \geq \bo, \sum_i c_i = 1}{\norm{\yob - \lp \bA\bG\rp\bc}_2^2}
\end{equation}
where the columns of the matrix $\bG$, $\lc \bg_i \rc$, represent an element of the dictionary and $\bc$ is the vector of coefficients respectively assigned to these elements. The first approach above has the advantage of significantly reducing the effective dimensionality of the problem \cite{raj2014multi}, but requires the solution of the highly nonlinear problem of determining the $2M$ variables $\sigma_i$ and $\mu_i$. On the other hand, the second approach requires only sufficient knowledge of the system under study to permit a reasonable selection of dictionary functions; nonlinearity is introduced through the positivity constraint. This problem becomes increasingly ill-posed with increasing $M$, and will in general be severely ill-posed due both to the required value of $M$ and the non-orthogonality of the Gaussian dictionary functions.
\section{The Span of Regularization Method}
\label{sec_Multi-Reg}
\subsection{Motivation}
To motivate SpanReg, we first discuss the artificial case in which $\f\geq \bo$ is known, from which a noise-corrupted signal $\yob = \bA \f + \bomega$ can be created; we now determine $\f_{\lambda_j}$ for specific values of $\lambda_{j}$ within a set of regularization parameters $\Lambda$ by solving Eq. \eqref{constrained_tikhonov_regularization}. 

As indicated above, the various available methods for selection of $\lambda$ in effect evaluate results for several $\lambda$'s and retain only the one corresponding to an optimal value, discarding the others.  
Thus, we define our approximation of $\f$ as a linear combination of regularized solutions:  
\begin{equation}
	\f \approx \sum_{j=1}^N \alpha_j \f_{\lambda_j}
\end{equation}
where $\left\lbrace \alpha_j \right\rbrace_1^N $ is the solution to the least squares problem
\begin{equation}
	\alpha=\left(\alpha_1,\alpha_2,\cdots,\alpha_N \right) = \argmin_{\alpha \geq \bo} \norm{\f - \sum_{j=1}^N \alpha_j \f_{\lambda_j}}_2^2
	\label{offline_alpha}.
\end{equation}
When $\f$ is known, the set of $\alpha_j$'s defined in Eq. \eqref{offline_alpha}, with each $\alpha_j$ corresponding to an element of a linear combination of regularized solutions, can always be selected to provide an improvement in signal reconstruction as compared to selection of any single value of lambda. This motivates the SpanReg approach for reconstructing an unknown distribution. However, an algorithm for selecting the best set $\left\lbrace \alpha_j \right\rbrace_{{j=}1}^N$ in the approximation:
\begin{equation}
	\f \approx \sum_{j=1}^N \alpha_j \f_{\lambda_j} =: \f_{\alpha},\quad \text{where } \alpha_j \geq 0 
	\label{f_unknown_alpha}.
\end{equation}
for an unknown DF, $\f$, must now be developed. 

An additional relationship arises from the representation of $\f$ as a linear combination of Gaussian dictionary functions $\left \lbrace\bg_i \right \rbrace_{{i=}1}^M$, which we impose with a non-negativity constraint on the expansion coefficients:  
\begin{equation}
	\f \approx \sum_{i = 1}^{M} c_i\bg_i =: \f_{\bc},\quad \text{where } c_i \geq 0.
	\label{f_unknown_c}
\end{equation}
We consider the case in which the integral over $\f$ equals $1$, so that it represents a probability distribution function (PDF). Since $\int g\lp \sigma_i, \mu_i\rp = 1$, we have the requirement that $\sum c_i = 1$.

\subsection{Analytic Framework}
\label{sec: multireg}
\subsubsection{Theory}
To link the two expressions (\ref{f_unknown_alpha}) and (\ref{f_unknown_c}), we define a symbolic inversion operator $\bA^{-1}_{\lambda_j}$ as follows: given a vector of noisy observations $\by_{\text{ob}}$, $\bA^{-1}_{\lambda_j}\by_{\text{ob}}:=\f_{\lambda_j}$, {  where $\f_{\lambda_j}$ is the solution of \eqref{constrained_tikhonov_regularization} with $\lambda=\lambda_j$, $j=1,\cdots,N$.} This operator is not the usual matrix pseudoinverse because of the non-negativity constraints.

We apply this to the noisy observations that would correspond to the signal generated by the $\bg_i$, obtaining a set of associated DF's $\{\bg_{i,\lambda_j}\}${, $i=1,\cdots,M$, $j=1,\cdots,N$} according to Eq. \eqref{constrained_tikhonov_regularization}:
\begin{equation}
    \bg_{i,\lambda_j} =  \bA^{-1}_{\lambda_j} \left( \bA \bg_i + \bomega\right).
    \label{time_domain_simulation}
\end{equation}
where $\bomega$ represents a noise realization exhibiting the same noise level, defined by RMS($\omega$), as the observed data $\by_{\text{ob}}$. Note that while determining noise amplitude within an experimental signal can be highly problematic, in our examples of decaying exponentials this information is directly available from data collected from the signal tail.

We now seek to approximate both $\f_{\alpha}$ and $\f_{\bc}$ by the $\{\bg_{i,\lambda_j}\}$. 
We first write 
\begin{equation}
\bg_i  \approx\sum_{j = 1}^N \beta_{ij} \bg_{i,\lambda_j}, \quad \beta_{ij}\geq 0.
\label{eqn_for_beta}
\end{equation}
The expansion coefficients $\beta_{ij}$ would be Kronecker $\delta$'s in the absence of noise and regularization.
As $\bomega$ is random, the corresponding $\{\bg_{i,\lambda_j}\}$ and $\{\beta_{ij}\}$ are random variables. Let $\langle \cdot \rangle$ denote the ensemble average of a random variable over $n_{\text{run}}$ noise realizations. In particular,
\begin{equation*}
    \langle\bg_{i,\lambda_j}\rangle = \frac{1}{n_{\text{run}}} \sum_{k = 1}^{n_{\text{run}}} \bg_{i,\lambda_j}^{(k)},\quad \langle\beta_{ij}\rangle = \frac{1}{n_{\text{run}}} \sum_{k = 1}^{n_{\text{run}}} \beta_{ij}^{(k)}
\end{equation*}
where the superscript $(k)$  indicates the $k$-th noise realization. 
We now apply a representation analogous to Eq. \eqref{f_unknown_c} to define regularized approximations:
\begin{equation}
\f_{\lambda_j} \approx  \sum_{i = 1}^M x_{ij} \langle\bg_{i,\lambda_j} \rangle
\label{eqn_for_x}
\end{equation}
where the $\{x_{ij}\}$ can be obtained through least squares analysis.  

Thus, given the dictionary of Gaussian distributions $\left \lbrace \bg_{i}\right \rbrace_{i=1}^M$, we can find the coefficients $\left \lbrace \langle\beta_{ij}\rangle\right \rbrace$ and $\left \lbrace x_{ij}\right \rbrace$ from the approximations to the regularized solutions $\left \lbrace \f_{\lambda_j} \right \rbrace$ and $\left \lbrace \langle\bg_{i,\lambda_j}\rangle\right \rbrace$, respectively. Now equating (\ref{f_unknown_alpha}) and (\ref{f_unknown_c}) and using (\ref{eqn_for_beta}) and (\ref{eqn_for_x}), we arrive at an expression containing only $\{\langle\beta_{ij}\rangle\}$, $\{x_{ij}\}$ and $\{\langle\bg_{i,\lambda_j}\rangle\}$:
\begin{equation}
	\begin{aligned}
	\sum_{j = 1}^N \alpha_j \f_{\lambda_j} &\approx \sum_{i = 1}^M c_i \bg_i\\
	\Rightarrow \f_{\alpha} := \sum_{j = 1}^N \alpha_j \sum_{i = 1}^M x_{ij} \langle\bg_{i,\lambda_j}\rangle&\approx \sum_{i = 1}^M c_i \sum_{j = 1}^N \langle\beta_{ij}\rangle \langle\bg_{i,\lambda_j}\rangle =: \f_{\bc}
	\end{aligned}
\end{equation}
The coefficients $\boldsymbol{\alpha}$ and $\bc$ are obtained by solving the least squares problem:
\begin{equation}
\left \{ 
\begin{aligned}
(\boldsymbol{\alpha}^*,\mathbf{c}^*) = \argmin \norm{\f_{\alpha} - \f_{\bc}}_2\\
\text{subject to } \boldsymbol{\alpha^*} \geq \bo,\; \mathbf{c}^*\geq \bo, \text{ and }\sum_i c^*_i = 1, 
\end{aligned}\right.
\label{min_l2}
\end{equation}
Our expression for the final recovered $\f^*$ is then:
\begin{equation}
	\f^*_{\alpha} = \sum_{j=1}^N \alpha_j{^*} \f_{\lambda_j}
	\label{final_rep}
\end{equation}
Eq. \ref{final_rep} is the main result of our analysis, defining the desired recovered DF in terms of a linear combination of differently-regularized solutions based on the observed data. Alternatively, the corresponding result for $\f^*_c$ may also be used with essentially equivalent results.

\subsubsection{Numerical Implementation}
We define the following notation:	
\begin{align*}
  \bL_{\alpha} &= \begin{pmatrix}
\begin{pmatrix}
\langle\bg_{1,\lambda_1}\rangle& \langle\bg_{2,\lambda_1}\rangle & \cdots & \langle\bg_{M,\lambda_1}\rangle 
\end{pmatrix}
&
,\cdots,
&
\begin{pmatrix}
\langle\bg_{1,\lambda_N}\rangle & \langle\bg_{2,\lambda_N}\rangle  & \cdots & \langle\bg_{M,\lambda_N}  \rangle
\end{pmatrix}
\end{pmatrix} \in \R^{n\times MN}\\
 \bL_{c} &= \begin{pmatrix}
\begin{pmatrix}
\langle\bg_{1,\lambda_1}\rangle & \langle\bg_{1,\lambda_2}\rangle  & \cdots & \langle\bg_{1,\lambda_N}\rangle  
\end{pmatrix}
&
,\cdots,
&
\begin{pmatrix}
\langle\bg_{M,\lambda_1}\rangle & \langle\bg_{M,\lambda_2}\rangle  & \cdots & \langle\bg_{M,\lambda_N}\rangle  
\end{pmatrix}
\end{pmatrix}\in \R^{n\times MN}\\
\bx_{\text{vec}}  &= \begin{pmatrix}
\begin{pmatrix}
x_{11} & x_{21}  & \cdots & x_{M1} 
\end{pmatrix}
&
,\cdots,
&
\begin{pmatrix}
x_{1N}  & x_{2N}  & \cdots & x_{MN}  
\end{pmatrix}
\end{pmatrix}\in \R^{MN}\\
\boldsymbol{\beta}_{\text{vec}}  &= \begin{pmatrix}
\begin{pmatrix}
\langle\beta_{1,\lambda_1} \rangle& \langle\beta_{1,\lambda_2}\rangle  & \cdots & \langle\beta_{1,\lambda_N}\rangle 
\end{pmatrix}
&
,\cdots,
&
\begin{pmatrix}
\langle\beta_{M,\lambda_1}\rangle & \langle\beta_{M,\lambda_2}\rangle  & \cdots & \langle\beta_{M,\lambda_N}\rangle  
\end{pmatrix}
\end{pmatrix}\in \R^{MN}
\end{align*}
where {for} each {pair of fixed $i$ and $j$, the } regularized approximation $\langle\bg_{i,\lambda_j}\rangle$ is a column vector of dimension $\R^n$ and the elements $x_{ij}$ and $\langle\beta_{i\lambda_j}\rangle$ are scalars.  Thus, the lengths of the vectors $\bx_{\text{vec}} $ and $\boldsymbol{\beta}_{\text{vec}} $ are equal to the number of columns in $\bL_{\alpha}$ and $ \bL_{c}$, respectively. Moreover, we write for the unknowns:
\begin{equation}\nonumber
\boldsymbol{\alpha}_{\text{vec}}  = \begin{pmatrix}
\begin{pmatrix}
\alpha_1 & \alpha_1  & \cdots &\alpha_1  & \alpha_1 
\end{pmatrix}_{M}
&
,\cdots,
&
\begin{pmatrix}
\alpha_N  & \alpha_N  & \cdots & \alpha_N  & \alpha_N 
\end{pmatrix}_{M}
\end{pmatrix}^T\in \R^{MN}
\end{equation}
\begin{equation}\nonumber
\mathbf{c}_{\text{vec}}  = \begin{pmatrix}
\begin{pmatrix}
c_1 & c_1  & \cdots & c_1  & c_1 
\end{pmatrix}_{N}
&
,\cdots,
&
\begin{pmatrix}
c_M & c_M  & \cdots & c_M  & c_M 
\end{pmatrix}_{N}
\end{pmatrix}^T \in \R^{MN}
\end{equation}
Then Eqs. \eqref{f_unknown_alpha} and \eqref{f_unknown_c} can be written:
\begin{equation}\nonumber
\f_{\alpha} =    \bL_{\alpha}\cdot \text{diag}\left(\bx_{\text{vec}}   \right)\cdot \boldsymbol{\alpha}_{\text{vec}} 
\end{equation}
\begin{equation}\nonumber
\f_{c} =  \bL_{c}\cdot \text{diag}\left(\boldsymbol{\beta}_{\text{vec}}  \right)\cdot \mathbf{c}_{\text{vec}} 
\end{equation}
where $\cdot$ denotes the usual matrix-vector multiplication and $\text{diag}$ indicates the diagonal matrix formed from the vector argument. Eq. \eqref{min_l2} can then be re-formulated as: 
\begin{equation}
\left \{ \begin{aligned}
(\mathbf{c}^*,\boldsymbol{\alpha}^*) &= \argmin \norm{   \bL_{\alpha}\cdot \text{diag}\left(\bx_{\text{vec}}   \right)\cdot \boldsymbol{\alpha}_{\text{vec}}  -  \bL_{c}\cdot \text{diag}\left(\boldsymbol{\beta}_{\text{vec}}  \right)\cdot \mathbf{c}_{\text{vec}} }_2\\
&\text{subject to } \bc \geq \bo,\boldsymbol{\alpha}  \geq \bo, \sum_i c_i  = 1
\end{aligned}\right.
\label{offline_implicit_form_ls}
\end{equation}
By writing the solution in the stacked form: \begin{equation}
\bs = \lp\alpha_1,\alpha_2,\cdots,\alpha_{N},c_1,c_2,\cdots,c_M \rp^T \in \R^{N+M,}
\end{equation}
\eqref{offline_implicit_form_ls} can be expressed as a conventional LS problem for $\bs^* \in \R^{N+ M}$
\begin{equation}
\left \{ \begin{aligned} 
\bs^* &= \argmin_{\bs \geq \mathbf{0}} \norm{\bB \bs}_2\\
&\text{such that } \sum_{j=N+1}^{N+M} s_j = 1
\end{aligned}\right.
\label{online_explicit_form_ls}
\end{equation}
with 
\begin{equation}
\bB =    \bL_{\alpha}\cdot \text{diag}\left(\bx_{\text{vec}}   \right)\cdot  \mathbf{TT}_{\alpha} -  \bL_{c}\cdot \text{diag}\left(\boldsymbol{\beta}_{\text{vec}}  \right) \cdot \mathbf{TT}_{c} \in \R^{n\times (N+M)}
\end{equation}
\begin{equation}
\mathbf{TT}_{\alpha} = \bI_N \otimes \begin{pmatrix}
1\\ 
1\\ 
\vdots\\ 
1
\end{pmatrix}_M \begin{pmatrix}
\bI_{N} &
\mathbf{0}_{ N \times M}
\end{pmatrix} \in \R^{MN \times (N+M)}
\end{equation}
and
\begin{equation}
\mathbf{TT}_{c} = \bI_M \otimes \begin{pmatrix}
1\\ 
1\\ 
\vdots\\ 
1
\end{pmatrix}_N\begin{pmatrix}
\mathbf{0}_{M \times N} &
\bI_{M}
\end{pmatrix} \in \R^{MN \times (N+M)}
\end{equation}
where $\otimes$ is the Kronecker tensor product, $\bI_M$, $\bI_N$ are {the} identity matrices of { rank} $M$ and $N$, respectively, and $\bo_{N\times M}$ is the zero matrix of size $N\times M$.

The computational procedure can be divided into two parts, {which we refer to as the }offline {part}, meaning independent of the actual data set, and {the }data-dependent online {part}. In the offline computation, $\left\lbrace \bg_i \right\rbrace $, $\left\lbrace \langle\bg_{i,\lambda_j}\rangle \right\rbrace $ and $\left\lbrace \langle\beta_{i,\lambda_j}\rangle \right\rbrace $ are determined only once. For the online part, for each noisy measurement $\by_{\text{ob}}$, the corresponding $\left\lbrace \f_{\lambda_j}\right\rbrace $, $\left\lbrace x_{ij}\right\rbrace $ and $(\mathbf{c}^*,\boldsymbol{\alpha}^*) $ can be obtained with the desired solution given by (\ref{final_rep}). { Note} that the {$\bg_{i,\lambda_j}$}, and hence the {$\beta_{i,\lambda_j}$} {in (\ref{eqn_for_beta})}, are noise-dependent.
For computations, we  use the ensemble average over $n_{\text{run}}$ of realizations of Eq. (\ref{eqn_for_beta}).

The pseudocode for the offline and online computations reads as follows:\\
\begin{algorithm}[H]
	\SetAlgoLined
	\KwData{Pre-determined $\left\lbrace \lambda_j\right\rbrace$, $\left\lbrace \bg_i \right\rbrace$, $n_{\text{run}}$ and noise SNR}
	\KwResult{$\left\lbrace \langle\bg_{i,\lambda_j}\rangle \right\rbrace $ and $\left\lbrace \langle\beta_{i,\lambda_j}\rangle \right\rbrace $}
	initialization\;
	\For{$k= 1,2,\cdots,n_{\text{run}}$,}{generate noise realization $\omega_k$
		\For{$i = 1,2,\cdots, M$}{
		$\by_{\text{clean}}^{(i)} = A\bg_i$\;
		$\by_{\text{noisy}}^{(i)}(\omega_k) = \by_{\text{clean}}^{(i)} + \omega_k$\;
		\For{$j = 1,2,\cdots, N$}{
			$\bg_{i,\lambda_j}(\omega_k) = \argmin_{\bg \geq \bo} \left\lbrace \norm{\bA \bg_i - \by_{\text{noisy}}^{(i)}(\omega_k)}_2^2 + \lambda_j^2\norm{\bg_i}_2^2  \right\rbrace $\;
		}
		}
	$\beta_{ij}(\omega_k) = \argmin_{b_{ij} \geq 0}\norm{\bg_i - \sum_{j = 1}^N b_{ij} \bg_{i,\lambda_j}(\omega_k)}_2$}
	$\langle\bg_{i,\lambda_j}\rangle = \frac{1}{n_{\text{run}}}\sum_{k = 1}^{n_{\text{run}}} \bg_{i,\lambda_j}(\omega_k)$ and $\langle\beta_{i,\lambda_j}\rangle= \frac{1}{n_{\text{run}}}\sum_{k = 1}^{n_{\text{run}}} \beta_{ij}(\omega_k)$
	\caption{Offline Computation}
\end{algorithm}

\begin{algorithm}[H]
	\SetAlgoLined
	\KwData{$\by_{\text{ob}}$, $\left\lbrace \lambda_j\right\rbrace$, $\left\lbrace \langle\bg_{i,\lambda_j}\rangle\right\rbrace $, $\left\lbrace \langle\beta_{i,\lambda_j}\rangle \right\rbrace $}
	\KwResult{$\f^*$}
	initialization\;
	\For{$j = 1,2,\cdots,N$}{
	$\f_{\lambda_j} = \argmin_{\fhat \geq \bo} \left\lbrace \norm{\bA \fhat  - \yob}_2^2 + \lambda_j^2\norm{\fhat }_2^2  \right\rbrace $\;
	$\bx_{\cdot,j} = \argmin_{x_{ij} \geq 0}\norm{\f_{\lambda_j} - \sum_{i = 1}^M x_{ij} \langle \bg_{i,\lambda_j}\rangle}_2$\;
	}
	$(\mathbf{c}^*,\boldsymbol{\alpha}^*) = \argmin_{\mathbf{c}, \boldsymbol{\alpha} \geq 0, \sum_i c_i = 1} \norm{\sum_{j = 1}^N \alpha_j \sum_{i = 1}^M x_{ij} \langle\bg_{i,\lambda_j}\rangle- \sum_{i = 1}^M c_i \sum_{j = 1}^N \langle\beta_{i,\lambda_j}\rangle \langle\bg_{i,\lambda_j}\rangle}_2$\; 
	and $\f^*_{\alpha} = \sum_j \alpha^*_j \f_{\lambda_j}$\;
	\caption{Online Computation}
\end{algorithm}

\section{Applications of SpanReg to One-dimensional Magnetic Resonance Relaxometry}
We illustrate the application of SpanReg to the inverse problem of MRR for transverse relaxation in one dimension, whose discretized version is described in Section \ref{discrete_setup}. The objective is to approximate the distribution of transverse relaxation times $T_2$ within a sample from the signal $\mathbf{y}_{ob}$. In the context of our formalism, this distribution function takes the role of $\f$. $T_2$ values will be expressed in milliseconds (ms).
\subsection{Parameter Settings for the Implementation of SpanReg}
\label{sec: settings}
Several user-defined parameters must be selected before proceeding with the off-line computation. We first fix the values of $N$, $M$ and $n_{\text{run}}$, which are the number of regularization parameters, the number of Gaussian functions in the dictionary and the number of noise realizations, respectively. The values of $\lambda_j$, $j=1,\cdots,N$ and the means and standard deviations (SD) of the Gaussian functions $\lc\bg_i\rc_{i=1}^M$ in the dictionary must also be selected.

The regularization parameters are determined based on L-curve analysis, with particular attention to parameters that provide a wide range of distinct solutions to Eq. \eqref{constrained_tikhonov_regularization}. {In the following simulations,} we chose {$N$} = 16 {with $\lambda_1,\cdots, \lambda_N$} logarithmically spaced { over the interval} $[10^{-6}, 10^1]${.} The $T_2$ axis range is from 1 - 200 ms, discretized at 1 ms intervals, i.e. $\triangle T_2=1$. We chose a dictionary consisting of three families of Gaussians, each of which has its mean values equally spaced along the $T_2$ axis, and has a specified SD. The three families respectively consist of 160 members with SD$=2$ ms, 40 members with SD$=3$ ms, and 20 members with SD$=4$ ms; $M = 160 + 40 + 20= 220$.

\subsection{Comparison of SpanReg and Classical Parameter Selection Methods}
\label{sec: SpanReg_vs_dp}
\begin{figure}[H]
\centering
\captionsetup{width=.8\linewidth}
\includegraphics[width=0.9 \linewidth]{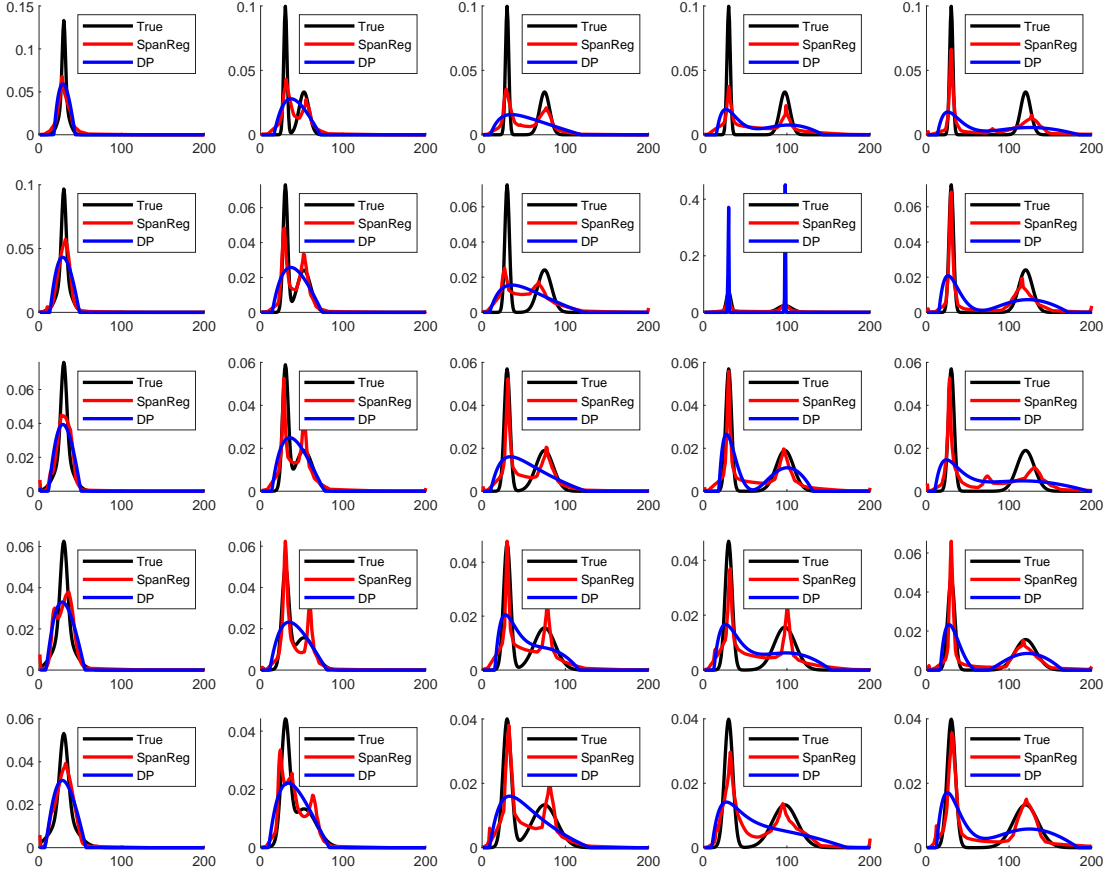}
\caption{\label{comp_two_peak_L}Comparison of {the} recovered distributions from SpanReg (red) and the discrepancy principle (blue) with the true underlying distributions (black). Each panel represents a $T_2$ distribution consisting of the sum of two Gaussian functions as in Eq. \eqref{simulation_form}. The value of $\sigma_1$, the standard deviation of the left-most component, increases from 2 ms (top row) to 5 ms (bottom row) with $\sigma_2=3\sigma_1$ throughout. Across each row, the ratio of peak separation (RPS) increases from 1 (left-most column) to 4 (right-most column), with the left-most distribution centered at $\mu = 35$ ms. }
\label{Recoveries_Comparision}
\end{figure}

We evaluated the performance of SpanReg for the reconstruction of a range of DFs consisting of two Gaussian functions with means $\lp\mu_1, \mu_2\rp$ and standard deviations $\lp \sigma_1, \sigma_2 \rp:$
\begin{equation}
    f_{\text{sim}}(T_2; \{\mu_1, \mu_2, \sigma_1, \sigma_2\}) = \frac{1}{\sqrt{2\pi \sigma_1^2}}e^{-\frac{(T_2 - \mu_1)^2}{2\sigma_1^2}} + \frac{1}{\sqrt{2\pi \sigma_2^2}} e^{-\frac{(T_2 - \mu_2)^2}{2\sigma_2^2}}.
    \label{simulation_form}
\end{equation}
We examine DF's with {$\sigma_1$ taking values in the set $\{2+0.75k,\, k=0,1,2,3,4\}$ (ms)} {and $\sigma_2=3\sigma_1$}. The separation between the Gaussians is defined by their ratio of peak separation (RPS), defined as RPS=$\frac{\mu_2}{\mu_1}$. We evaluate DF's with $RPS=1+0.75k$, $k=0,1,2,3,4$ for each $(\sigma_1, \sigma_2)$ pair. Note that none of the Gaussian functions in these distributions is included in the Gaussian dictionary $\lc \bg_i \rc$. We compare the recovery of each of the 25 resulting DF's (Fig. \ref{Recoveries_Comparision}) obtained from SpanReg and from the DP; here and elsewhere, by DP we indicate NNLS using the DP to set the Tikhonov regularization parameter. This is a particularly suitable comparison since for both methods, the noise level in the data must be known or estimated. As noted, the noise level can be estimated with good accuracy in the case of the superposition of exponentially decaying signals such as in an MRR experiment.  
The discretization along the $T_2$ axis is defined as above, with $n = 200$ evenly spaced values within the range $[1,200]$ ms. The measurement time vector is defined as $m=150$ sampling times evenly distributed in $t\in [0.3,400]$ ms. The additive noise level is set to result in SNR of $500$. The range of $\lambda$'s used for Tikhonov regularization
was from $10^{-6}$ to $10$ with {$N=16$} logarithmically spaced values. 

For the implementation of DP, the selected $\lambda_{DP}$ is determined by the following criterion \cite{hansen2010discrete}:
\begin{equation}
    \label{discrepancy_principle}
    \text{Choose }\lambda = \lambda_{DP} \text{ such that } \norm{\bA \f_{\lambda_{DP}} - \yob}_2 = \nu_{DP} \norm{\bomega}_2,
\end{equation}
where $\nu_{DP} \geq 1$ and $\norm{\bomega}_2 \approx \sqrt{m}\sigma(\bomega)$. We chose the safety factor $\nu_{DP}$ to be $\nu_{DP}=1.05$.  

Fig. \ref{Recoveries_Comparision} shows the comparison of  SpanReg with the DP for recovery of the indicated underlying DF for a single noise realization. As seen, SpanReg exhibits greater ability to resolve the two components of the DF and accurately model their amplitudes. 

\begin{figure}[H]
\captionsetup[subfigure]{labelformat=empty}
\centering
   \centering
   \subfloat[][]{\includegraphics[width=.45\linewidth]{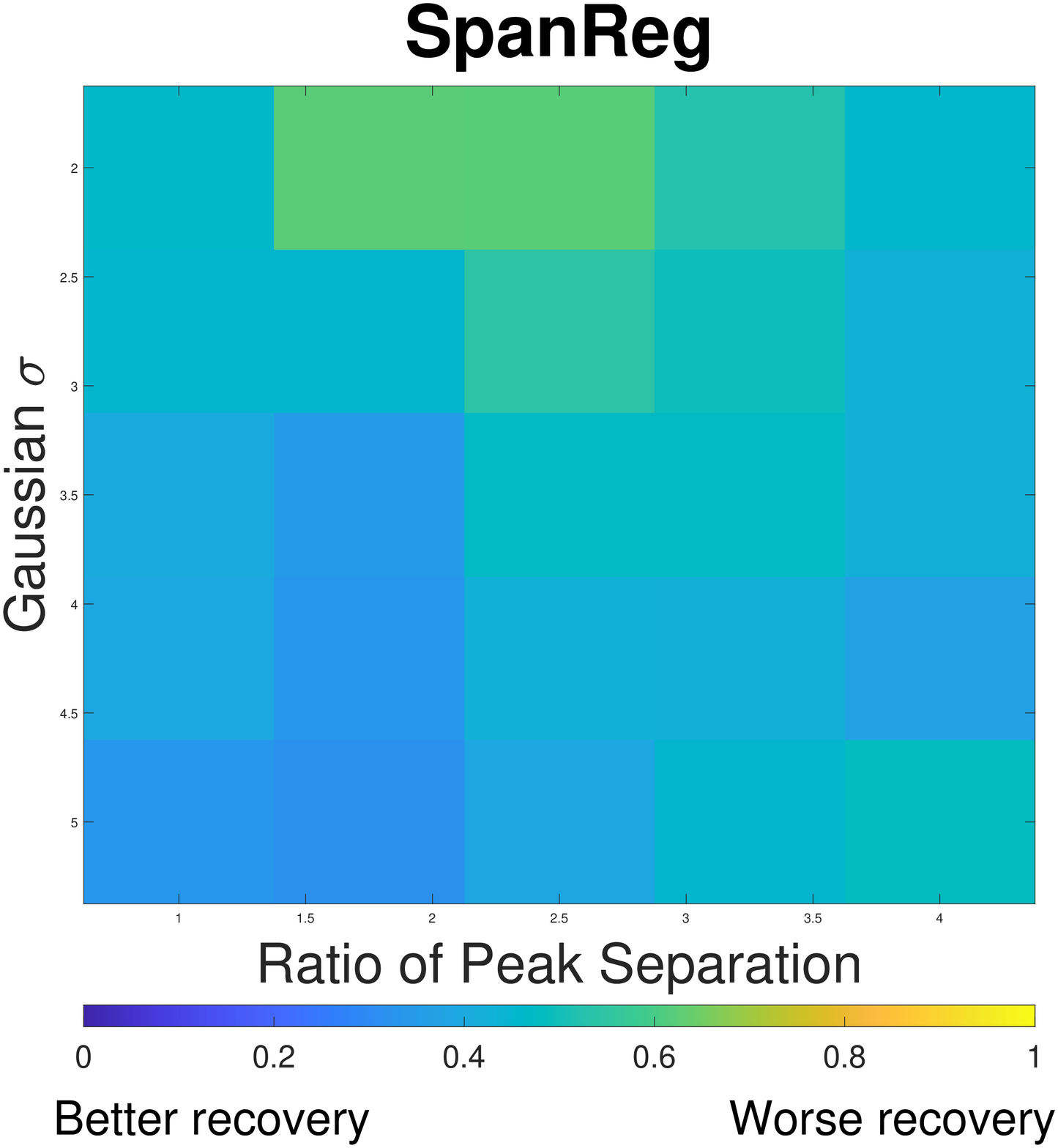}}\quad
   \subfloat[][]{\includegraphics[width=.45\linewidth]{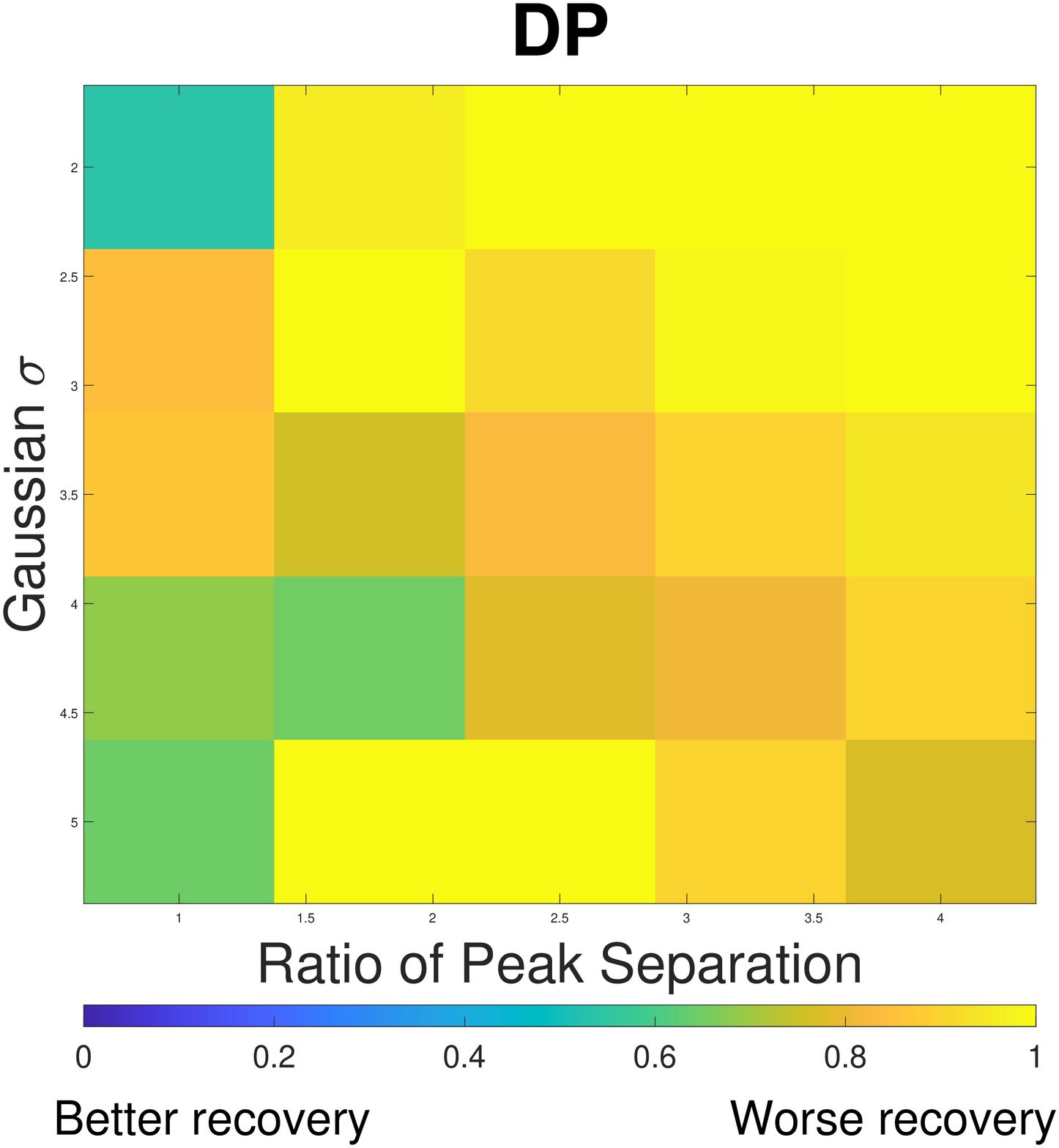}}\\
   \subfloat[][]{\includegraphics[width=.45\linewidth]{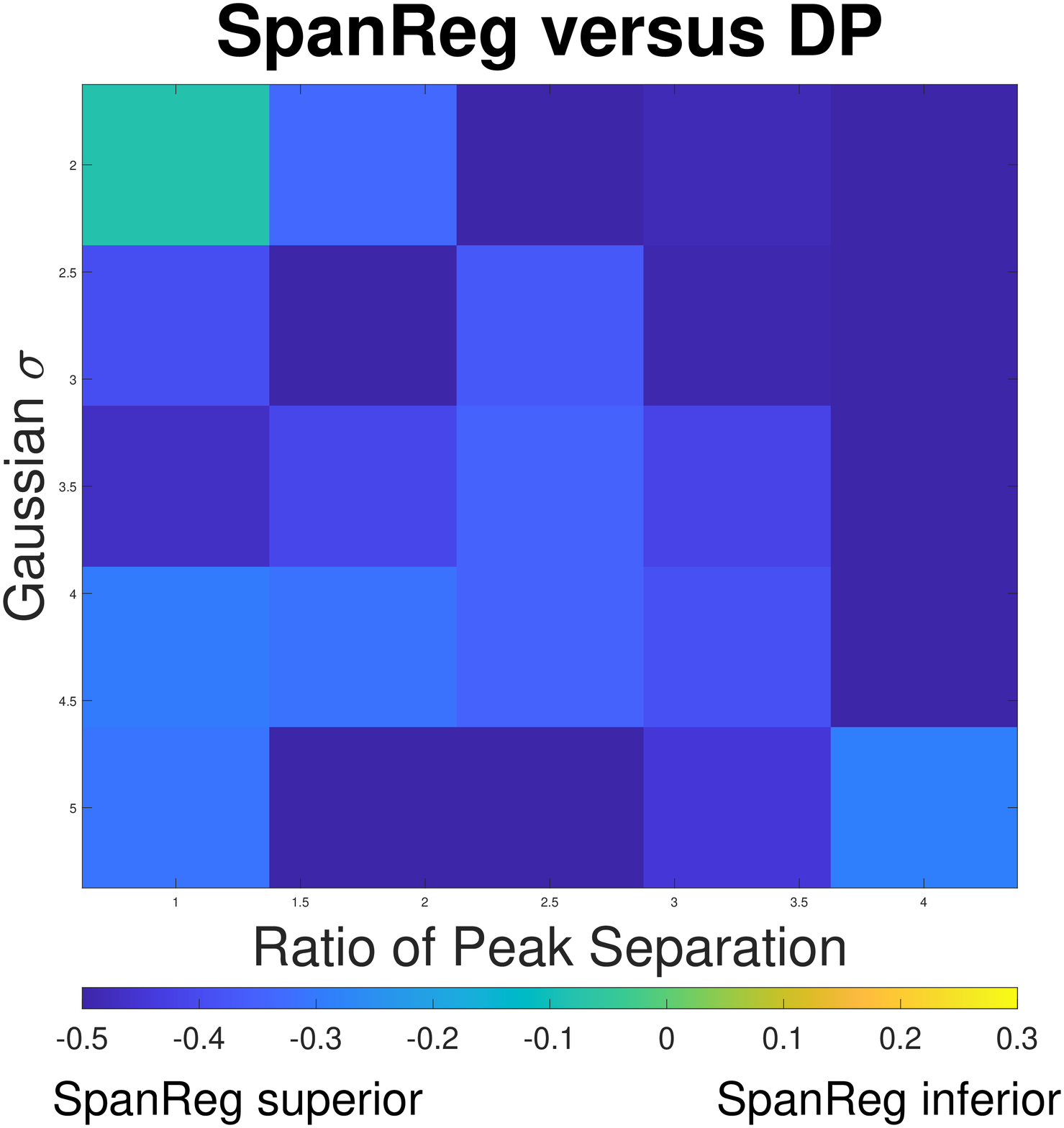}}
   \caption{Heat maps showing relative errors for SpanReg (upper-left), discrepancy principle (upper-right), and their differences $\varepsilon_{\text{diff}}$ (bottom), as defined in the text.}
   \label{comparisons_heatmap}
\end{figure}

Figure \ref{comparisons_heatmap} presents the comparison of the two reconstruction methods in terms of heat maps of mean relative error, defined respectively for SpanReg and the DP by
\begin{equation}
    \varepsilon_{\text{SpanReg}} = \frac{\norm{\f^*_{\text{SpanReg}} - \f_{\text{sim}}}_2}{\norm{\f_{\text{sim}}}_2},\, \varepsilon_{\text{DP}} = \frac{\norm{\f^*_{\text{DP}} - \f_{\text{sim}}}_2}{\norm{\f_{\text{sim}}}_2} \mbox{ and } \varepsilon_{\text{diff}} = \varepsilon_{\text{SpanReg}} - \varepsilon_{\text{DP}}
    \label{relative_error}
\end{equation}
where $\f^*$ is the reconstructed distribution and $\f_{\text{sim}}$ is defined in Eq. \eqref{simulation_form}. Thus, in terms of relative errors, a negative $\varepsilon_{\text{diff}}$ indicates the superiority of SpanReg reconstruction.
We see that by this summary metric of relative error, SpanReg outperforms DP across all DF's studied. Indeed, the errors for SpanReg are fairly constant across distribution, while the errors for DP increase as RPS increases over the illustrated range. Thus, as the component centers deviate from each other, including when closely-spaced, SpanReg is more able to accurately resolve them.  In addition, both methods tend to perform better as the widths of the Gaussian distributions $\sigma$ of the target DF increase; this highlights the smoothing characteristics of the $L_2$ norm. 

Additional insights are provided by comparing the heat map for the relative error metric described in Eq. \eqref{relative_error}, of use in summarizing the quality of reconstructions across a wide range of target DF{s}, with Fig. \ref{Recoveries_Comparision}. From the latter we see that for closely-spaced Gaussian components, represented by the second column of Fig. \ref{Recoveries_Comparision}, the reconstructions are not only quantitatively, but also qualitatively, different. The DP reconstruction incorrectly provides a single-component reconstruction in a number of cases, while SpanReg is clearly able to resolve the two underlying components.  Therefore, it is important to examine the actual recovered $T_2$ distributions in addition to comparing the single summary metric defined by relative error.

\begin{figure}[H]
\centering
\captionsetup{width=.8\linewidth}
\includegraphics[width=0.9 \linewidth]{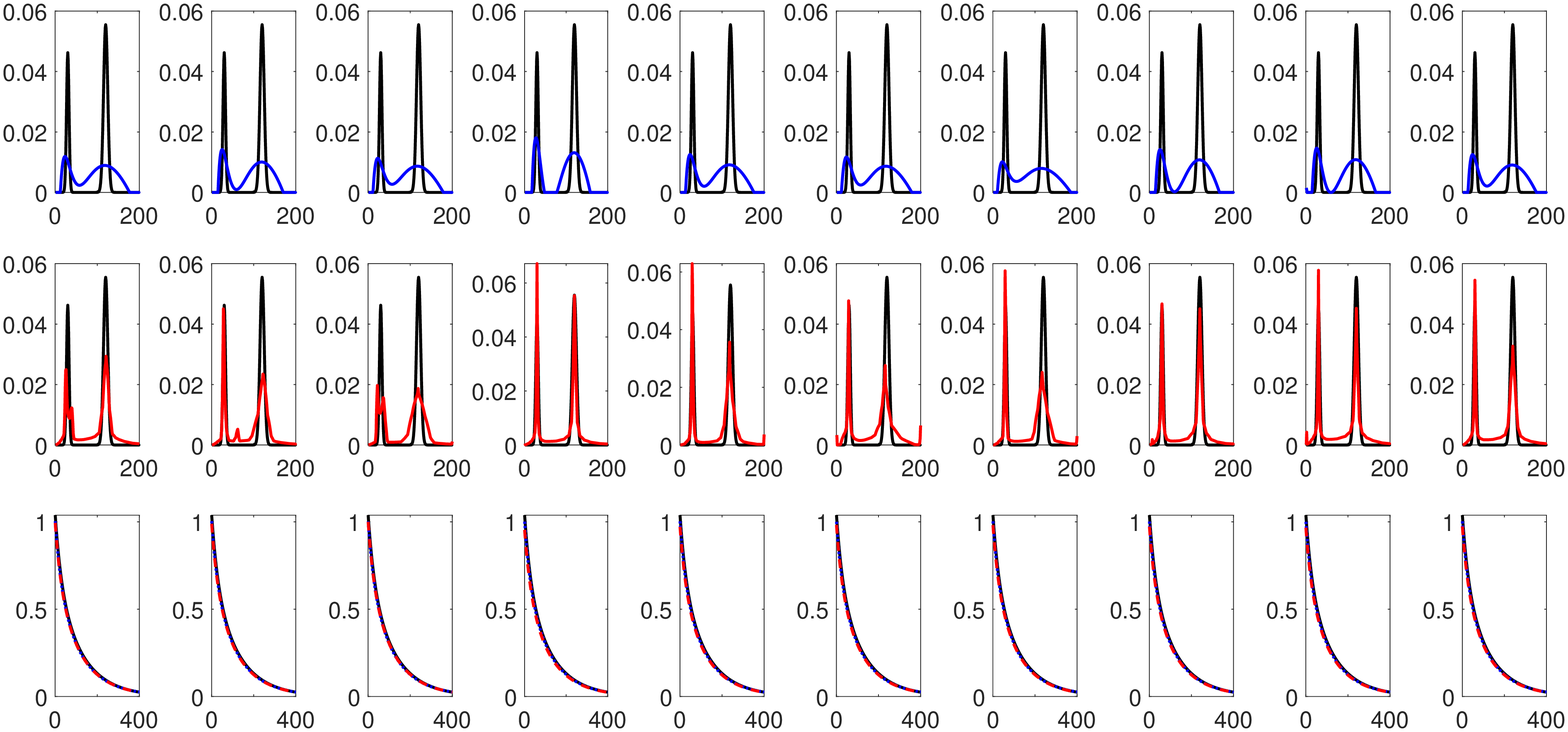}
\caption{Reconstruction of the sum of two Gaussians with $(\mu_1, \sigma_1$)=(30 ms, 3 ms) and $(\mu_2, \sigma_2)$=(120 ms, 5ms), shown in black in the top and middle rows; SNR=500. Each column {corresponds} to a single noise realization. Top: DP (blue), Middle: SpanReg (red), Bottom: signals generated from the DP solution and the SpanReg solution.}
\label{resolve_peaks_comparison_well_posed}
\end{figure}

\begin{figure}[H]
\centering
\captionsetup{width=.8\linewidth}
\includegraphics[width=0.9 \linewidth]{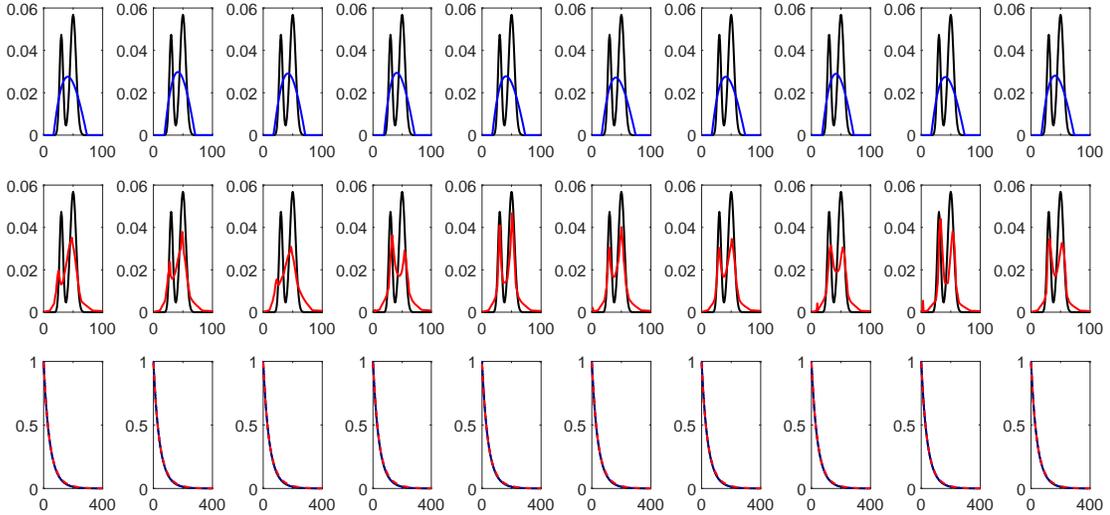}
\caption{Reconstruction of the sum of two Gaussians with $(\mu_1, \sigma_1$)=(30 ms, 3 ms) and $(\mu_2, \sigma_2)$=(50 ms, 5 ms), shown in black in the top and middle rows; SNR=500. Each column {corresponds} to a single noise realization. Top: DP (blue), Middle: SpanReg (red), Bottom: signals generated from the DP solution and the SpanReg solution.}  
\label{resolve_peaks_comparison}
\end{figure}

This is further highlighted in Figs. {\ref{resolve_peaks_comparison_well_posed}} and \ref{resolve_peaks_comparison}, comparing SpanReg and DP on two target distributions with respectively a greater and a lesser RPS. We show the results of reconstructions over $10$ noise realizations, as well as a comparisons of the corresponding signals generated by these reconstructions.
In Fig. \ref{resolve_peaks_comparison_well_posed}, SpanReg, as compared to the DP, provides a much more accurate reconstruction of the two components in the DF, although both methods accurately recover two resolved components.  In the much more ill-posed problem illustrated in Fig. \ref{resolve_peaks_comparison}, SpanReg clearly outperforms the DP in terms of stability with respect to noise and ability to resolve two closely-spaced peaks. DP fails to do this for all of the 10 noise realizations, while SpanReg succeeds in 7 out of 10 cases. Decreasing the DP safety factor improves the resolution performance of the DP, but introduces spurious peaks and greater noise sensitivity.

\subsection{Stability With Respect to Choice of Regularization Parameters}
\begin{figure}[H]
\captionsetup[subfigure]{labelformat=empty}
\centering
   \centering
   \subfloat[][]{\includegraphics[width=.45\linewidth]{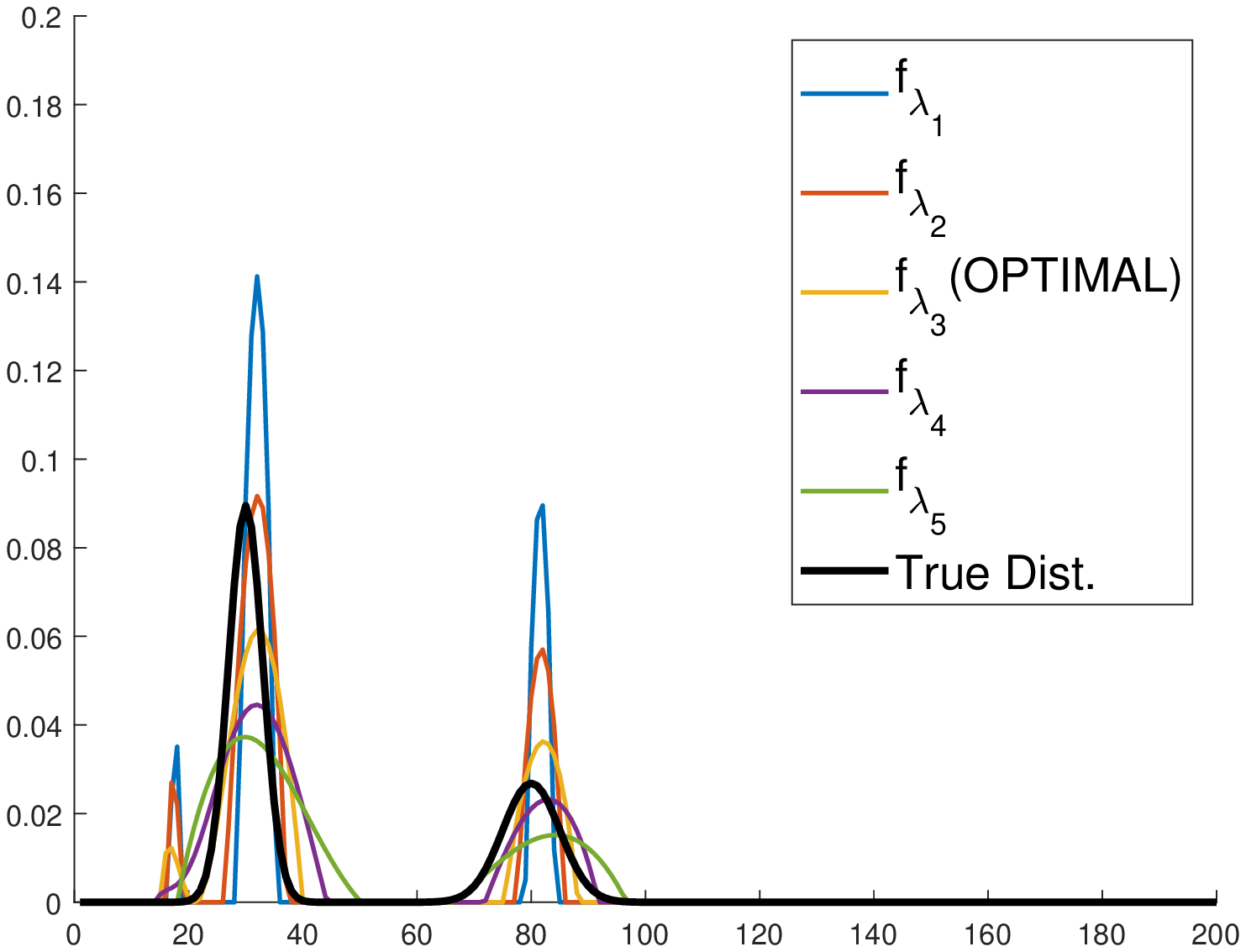}}\quad
   \subfloat[][]{\includegraphics[width=.45\linewidth]{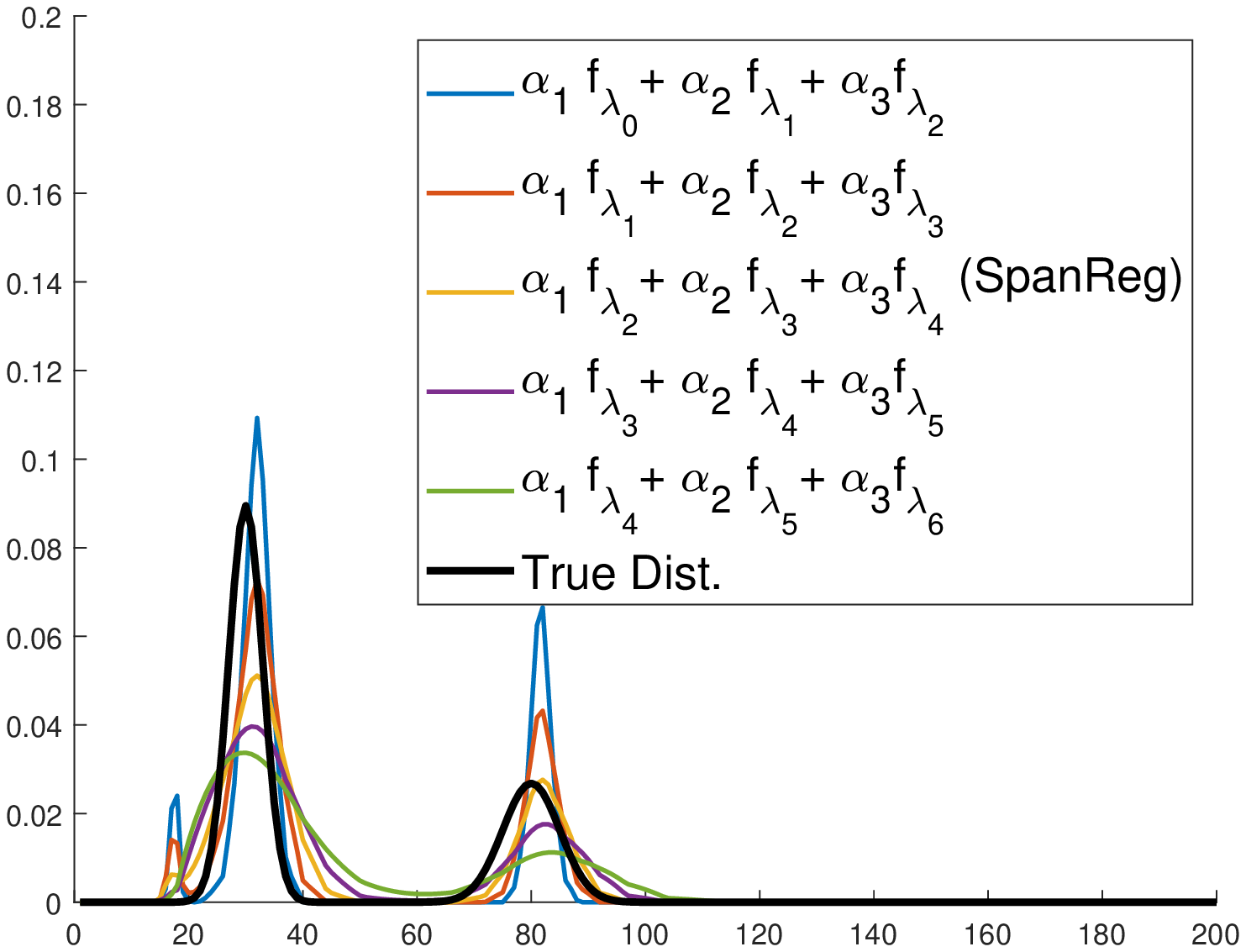}}\\
   \subfloat[][]{\includegraphics[width=.45\linewidth]{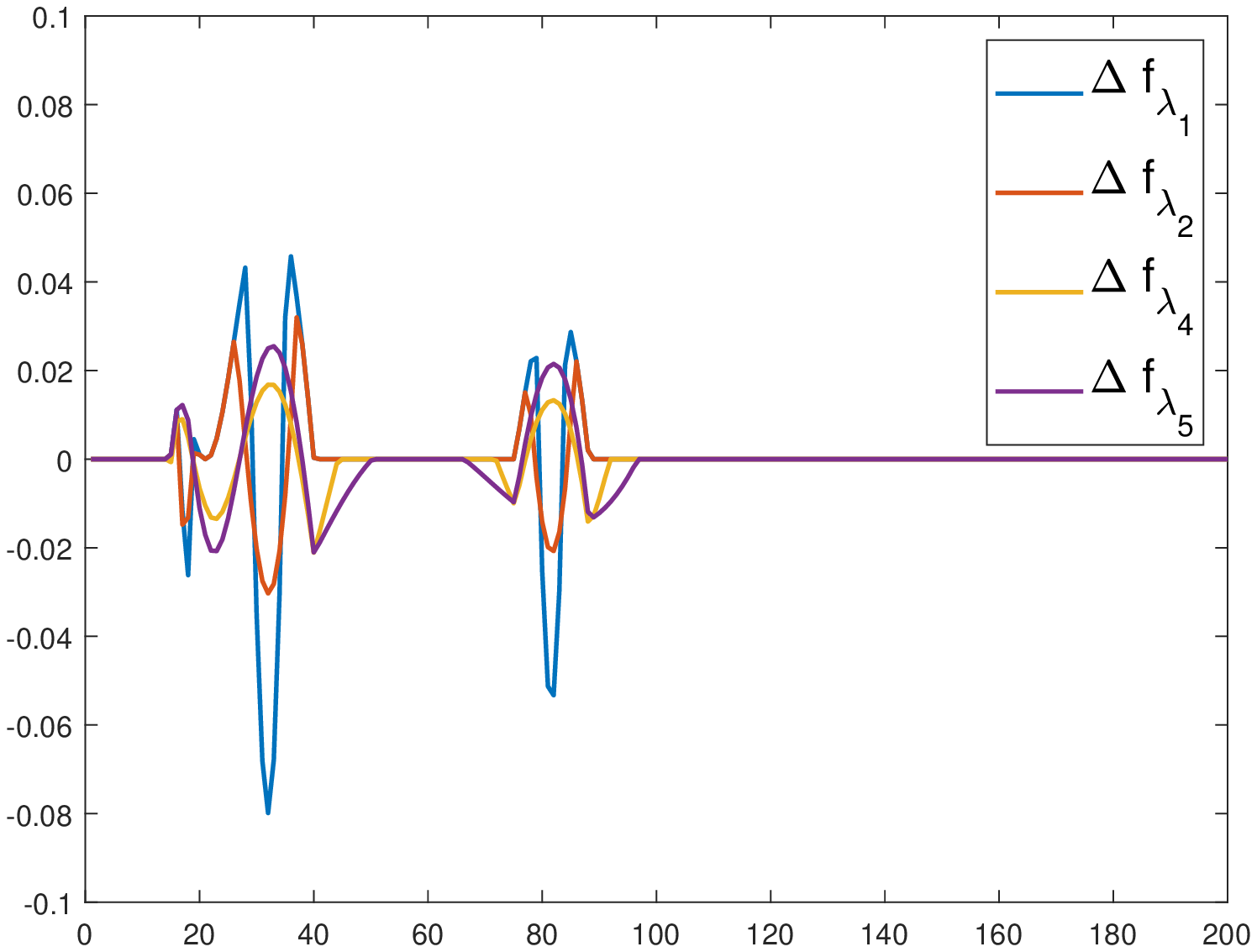}}\quad
   \subfloat[][]{\includegraphics[width=.45\linewidth]{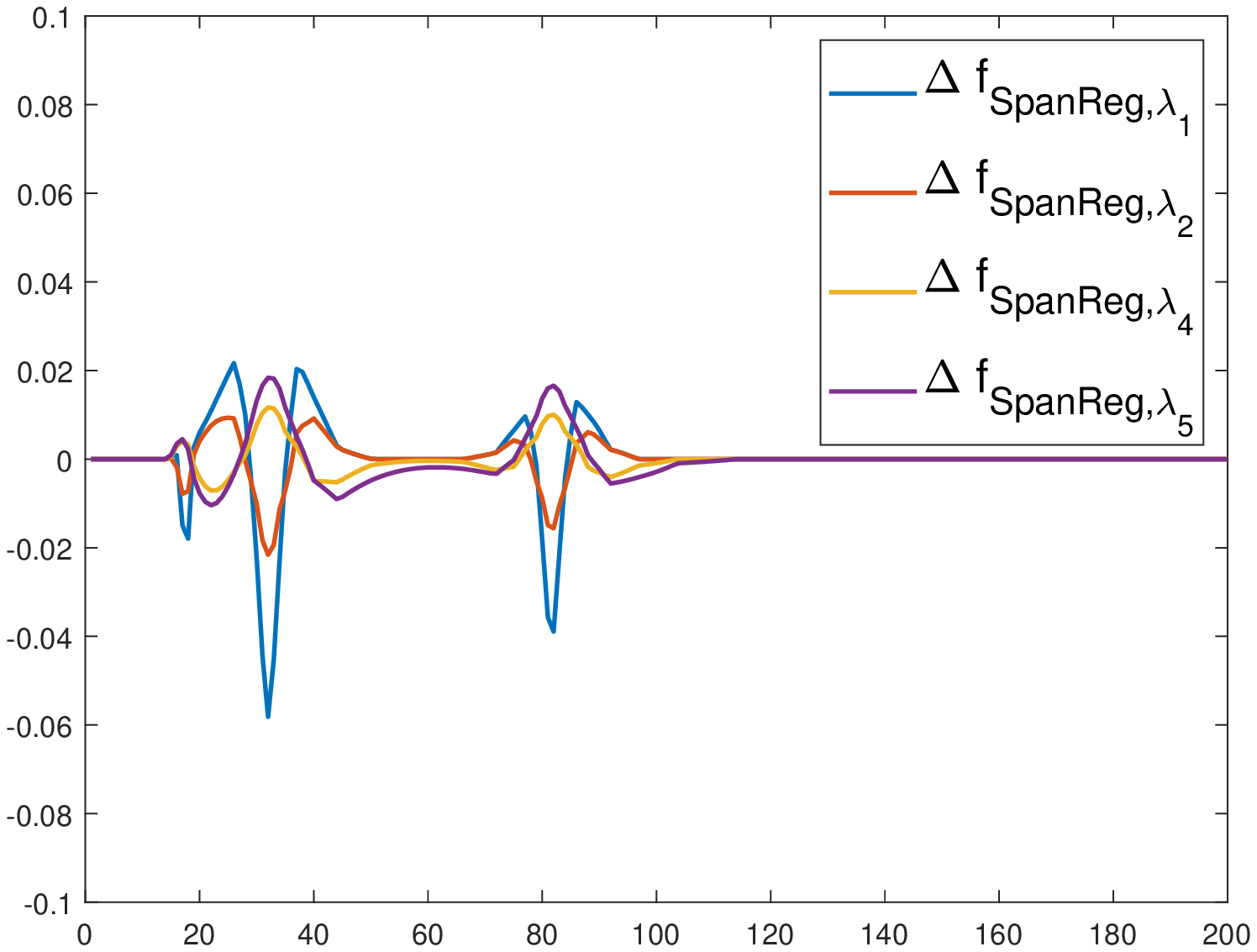}}\\   
   \caption{Stability with respect to choices of $\lambda$'s, using a single value of $\lambda$ for Tikhonov regularization and triplets of $\lambda$'s for SpanReg. In all cases, SNR=500. The true distribution is $\frac{2}{3}g(30,80)+\frac{1}{3} g(3,5)$. The optimal solution is $f_{\lambda_3}$. Upper left: Solutions obtained with optimal and sub-optimal regularizations. Upper right: Solutions obtained with SpanReg with subsequences of regularized solutions. Bottom left: Departures from optimal regularized solution in the upper left panel. Bottom right: Departures from corresponding SpanReg solution in the upper right panel.}
   \label{lambda_shift_stability}
\end{figure}

Selection of the optimal regularization parameter $\lambda$ is of critical importance in Tikhonov regularization, with deviations from that optimal value potentially resulting in substantial differences in the reconstructed DF. Fig. \ref{lambda_shift_stability} from the Supplementary Information shows the solutions obtained from Tikhonov regularization, and from SpanReg over different ranges of $\lambda$'s, including in all cases $\lambda_{\text{opt}}$.  In this simulation with a known underlying distribution, $\lambda_{\text{opt}}$ is determined by the $L_2$ error metric describing the difference between the underlying DF and the reconstructed DF:
\begin{equation}
    \label{lambda_opt}
    \lambda_{\text{opt}} = \argmin_{\lambda} \lc \norm{\f_{\lambda} - \f_{\text{sim}}}_2 \rc.
\end{equation}
{ Once $\lambda_{opt}$ is determined, stability is quantified by comparing the DF recovered using the different values of $\lambda$ from the set} $\mathcal{S}=\lc \frac{\lambda_{\text{opt}}}{2^4}, \cdots,\frac{\lambda_{\text{opt}}}{2^{1}},\frac{\lambda_{\text{opt}}}{2^0}, \frac{\lambda_{\text{opt}}}{2^{-1}},\cdots,\frac{\lambda_{\text{opt}}}{2^{-5}}\rc$. To apply this analysis to the DP, we simply compare solutions obtained with the different values of $\lambda$. We also show the differences between the results for these $\lambda$'s and the result for $\f_{\lambda_{\text{opt}}}$; larger values in these difference plots indicate less stability. For SpanReg, we perform the analysis using different triads of adjacent values of $\lambda$ from the { $\mathcal{S}$}. We form the reconstruction using the three $\lambda$'s centered around $\lambda_{\text{opt}}$, denoted $\f_{\text{SpanReg},\lambda_{\text{opt}}}$:
\begin{equation}
    \label{MR_f_lambda_opt}
    \f_{\text{SpanReg},\lambda_{\text{opt}}} := \alpha_1 \f_{\frac{\lambda_{\text{opt}}}{2}} + \alpha_2\f_{\lambda_{\text{opt}}} + \alpha_3 \f_{\frac{\lambda_{\text{opt}}}{2^{-1}}}, 
\end{equation}
where the values $\lc \alpha_1, \alpha_2,\alpha_3\rc$ are obtained from SpanReg. Then, to evaluate stability with respect to choice of $\lambda$, we plot results for SpanReg using three shifted values of $\lambda$'s. For example, for an integer $-3\leq j\leq 4$,
\begin{equation}
    \nonumber
    \f_{\text{SpanReg}, \frac{\lambda_{\text{opt}}}{2^j}} := \alpha_1 \f_{\text{SpanReg}, \frac{\lambda_{\text{opt}}}{2^{j+1}}} + \alpha_2\f_{\text{SpanReg}, \frac{\lambda_{\text{opt}}}{2^j}} + \alpha_3 \f_{\text{SpanReg}, \frac{\lambda_{\text{opt}}}{2^{j-1}}}.
\end{equation}
The comparison of the results for shifted triples of $\lambda$s defines the stability of SpanReg with respect to choice of regularization parameter. We also plot the difference between the DF recovered by adjacent shifted triplets to the the recovery obtained using the triplet centered on $\f_{\lambda_{\text{opt}}}$, i.e. we define $\Delta \f_{\frac{\lambda_{\text{opt}}}{2^j}} =\f_{\lambda_{\text{opt}}} -  \f_{\frac{\lambda_{\text{opt}}}{2^j}}$, and similarly, $\Delta \f_{\text{SpanReg},\frac{\lambda_{\text{opt}}}{2^j}} = \f_{\text{SpanReg},\lambda_{\text{opt}}} - \f_{\text{SpanReg},\frac{\lambda_{\text{opt}}}{2^j}}$. These plots quantify the changes in the recovered DF resulting from a suboptimal selection of $\lambda$ in SpanReg. A larger magnitude in these difference plots corresponds to lower stability with respect to choice of $\lambda$ and vice versa.

An example of this is shown in the Supplementary Information, Fig. \ref{lambda_shift_stability}, where results for the illustrated two-component Gaussian DF are shown for for optimally regularized DP solutions, and solutions obtained with sub-optimal regularization (left panels). Corresponding solutions obtained with SpanReg using different sub-sequences of regularized solutions are also shown (right panels). Departures from the optimal regularization solutions obtained for Tikhonov regularization (lower left panel) are substantially larger than those using SpanReg (lower right panel). The corresponding $L_2$ norms of the differences are $\lp \norm{\Delta \f_{\lambda_1}}, \norm{\Delta \f_{\lambda_2}}, \norm{\Delta \f_{\lambda_4}}, \norm{\Delta \f_{\lambda_5}} \rp = \lp 0.23, 0.11,    0.08,    0.13 \rp$ for Tikhonov regularization and\\ $\lp \norm{\Delta \f_{\text{SpanReg}, \lambda_1}}, \norm{\Delta \f_{\text{SpanReg}, \lambda_2}}, \norm{\Delta \f_{\text{SpanReg}, \lambda_4}}, \norm{\Delta \f_{\text{SpanReg}, \lambda_5}} \rp = \lp 0.14,    0.06,    0.05,    0.08 \rp$ for SpanReg, respectively, indicating a roughly 60$\%$ improvement in stability using SpanReg as compared to the DP.

\subsection{Myelin Water Fraction Mapping of the Human Brain}
Having shown the effectiveness of SpanReg on a wide range of simulated data, we now demonstrate its application to brain imaging. In particular, we compare the results of SpanReg, NNLS regularization using the DP, and non-regularized NNLS for conventional myelin water fraction (MWF) mapping in the human brain. In brief, this method involves acquiring a decaying bi-exponential signal for each image pixel, from which a DF of spin-lattice relaxation times $T_2$ is recovered. Given the high SNR of these images, we restrict our attention to the Gaussian noise approximation. Note again that the simulations above were targeted to parameter ranges appropriate for this problem of in vivo myelin mapping.  

3D gradient and spin-echo (GRASE) images were obtained from the brain of a healthy 49-year-old female using $32$ echoes at $TE_i= i\times TE$, where $TE = 11.3$ ms for $i = 1,2,\cdots, 32$. The notation $TE$, standing for echo time, is conventional in magnetic resonance, and corresponds to the measurement times in Eq. \eqref{laplace_transform_continuous}. The acquisition sequence had additional parameters of $TR = 1000$ ms, echo planar imaging acceleration factor of 3, field of view $278 \text{ mm} \times 200\text{ mm} \times 30\text{ mm}$, acquisition matrix of size $185 \times 133 \times 10$, acquisition voxel size $= 1.5 \text{ mm} \times 1.5\text{ mm} \times 3\text{ mm}$, reconstructed to $= 1\text{ mm} \times 1\text{ mm} \times 3\text{ mm}$ using zero filling in k-space. Scan time was approximately 10 minutes. A 3T Philips MRI system (Achieva, Best, the Netherlands), equipped with an internal quadrature body coil for transmission and an eight-channel phased-array head coil for reception, was used for acquisition.

We employ both unfiltered and NESMA-filtered datasets. NESMA \cite{Bouhrara_NESMA_2018} is a highly effective nonlocal denoising image filter that has been shown to increase the quality of myelin water fraction (MWF) mapping. The SNR of the unfiltered dataset is $\sim 10 - 300$, as defined by Eq. \eqref{SNR_def}, with variation across pixels due primarily to the variation in image amplitude.

While the ground truth for brain MWF in vivo is unknown, we can still implement numerical experiments using this data to compare different reconstruction approaches. First, we defined the MWF as the integral of the $T_2$ DF between the initial abscissa values of $T_2=6$ ms and $T_2= 40$ ms \cite{Bonny_2020,Bouhrara_NESMA_2018} for each pixel and created the corresponding map. To compare two different MWF maps $\bA$ and $\bB$, we define the scaled absolute difference (SAD) by:
\begin{equation}
    \label{scaled_absolute_difference}
    SAD(\bA, \bB) = \frac{S_{\bA - \bB}}{S_{\bA}}
\end{equation}
where $S_{\bM} = \sum_{i,j} \left\vert M_{i,j}\right\vert$ is the sum of the absolute values of the elements of a matrix $M_{i,j}$. In this way, if $SAD(\bA,\bB_1)$, as a comparison between a reference map $\bA$ and an approximation $\bB_1$, is smaller than $SAD(\bA,\bB_2)$, we say that the reconstructed MWF map $\bB_1$ exhibits superior accuracy to $\bB_2$. 

Because absolute magnetic resonance signal amplitudes are arbitrarily scaled by various sample characteristics, experimental settings and instrument properties, it is necessary to normalize the data so that $\int f(T_2) dT_2 = 1$ before applying SpanReg. This represents normalization of the PDF represented by the DF $f(T_2)$. Then, from Eq. \eqref{laplace_transform_continuous} written in terms of $TE$, the signal must be normalized to its initial value, that is $y(TE=0)=1$. This is problematic for two reasons. First, for experimental reasons, the spin-echo experiment in particular has measurement times at $i \times TE$ with $i=1$ rather than $i=0$. This limitation does not apply to other similarly-modeled experiments, such as diffusion-sensitizing pulse sequences. In addition, however, in any experiment, all data points are corrupted by noise so that normalization of the acquired signal by its initial value is not equivalent to normalizing the underlying signal, which is what is desired; the underlying and observed signals differ by noise at each data point. Instead, we proceed by setting the normalization factor to $\norm{\f_{LS}}_1$ according to the following:
\begin{equation}
    \begin{split}
        y_{LS}\lp TE = 0\rp &\approx \lp e^{-0/T_{2,1}}, e^{-0/T_{2,2}},\cdots,e^{-0/T_{2,n}}\rp \lp f_{LS,1}, f_{LS,2},\cdots,f_{LS,n} \rp^T \Delta T_2\\
        &= \sum_{j = 1}^n f_{LS,j} \Delta T_2 = \norm{\f_{LS}}_1 \Delta T_2 
    \end{split}\label{TE=0_formula}
\end{equation}
where the NNLS estimate of $\f_{LS}\geq \bo$ is determined from non-normalized data according to Eq. \eqref{NNLS}. 

We note that in effect, Eq. \eqref{NNLS} is equivalent to Eq. \eqref{fwd_eqn}, which is the original problem. In the current context, we are not representing the solution to Eq. \eqref{NNLS} to be an accurate estimate of the entire DF $\f$, but rather as the best available estimate of $f_{TE = 0}$. This approach introduces no additional bias though use of regularization. In our imaging analyses, the observed data $\yob$ is divided by $\norm{\f_{LS}}_1$ for each pixel. This analysis was not required for the simulation experiments above, where we illustrated SpanReg in the more usual case of signals with an initial abscissa value of zero. In Sections \ref{simulation_brain} and \ref{experiments_results}, we implement this approach to reconstruct SpanReg-based MWF maps.

We present two distinct analyses for the brain dataset.  First, in section \ref{simulation_brain}, we define a high-quality reference MWF map, based on pixel-wise reconstructions from SpanReg on NESMA-filtered data. This serves as a standard of comparison for three reconstruction methods, SpanReg and NNLS with DP regulariztion and without regulariztion, for two different levels of SNR.  Then, in the following section \ref{experiments_results}, we work directly with NESMA-filtered (high SNR) and non-filtered (low SNR) images using all three reconstruction methods to demonstrate their performance with respect to SNR; a comparison is made between results obtained at high and low SNR for each method separately.

\subsubsection{Comparison of Reconstruction Methods with Respect to a Reference Image}
\label{simulation_brain}
\begin{figure}[H]
\centering
\captionsetup[subfigure]{labelformat=empty}
\captionsetup{width=.8\linewidth}
\centering
\subfloat[]{}
{\includegraphics[width=0.29\linewidth]{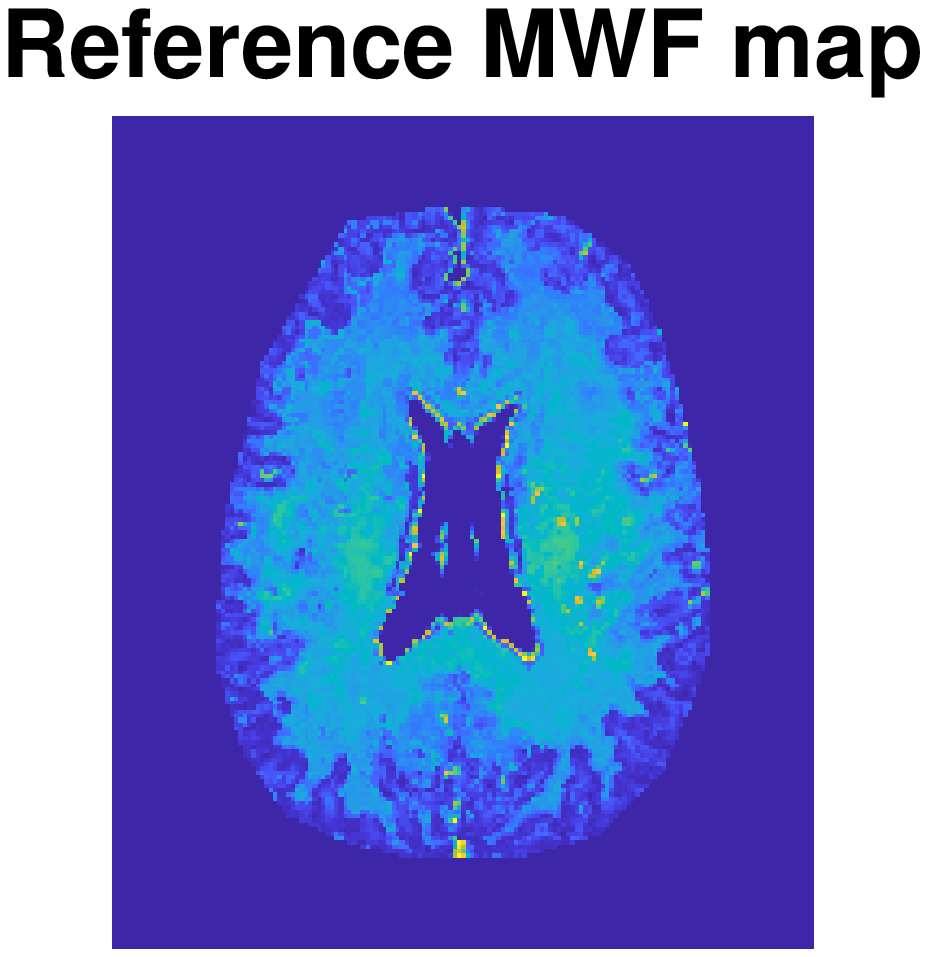}}
\subfloat[]{}{\includegraphics[width=0.9\linewidth]{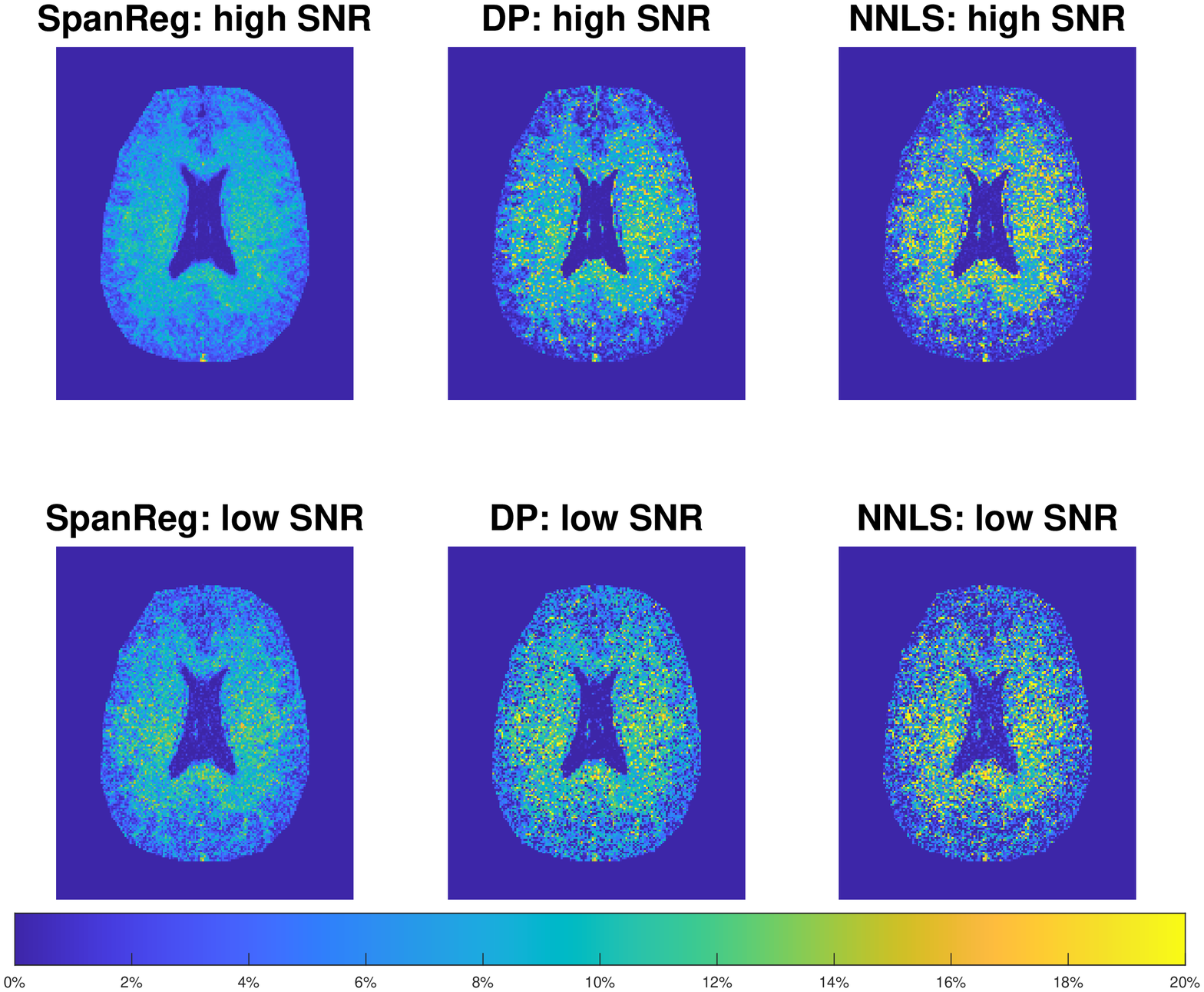}}
\caption{Uppermost image: reference MWF map; MWF maps generated from high SNR (top row; SNR=800) and low SNR (bottom row; SNR=200) data using SpanReg, NNLS with the DP, and non-regularized NNLS.}
\label{MWF_sim_GT}
\end{figure}

We obtained the reference MWF map shown in the uppermost panel of Fig. \ref{MWF_sim_GT} as follows. First, we obtained a stack of successively $T_2$-weighted images from the data acquisition sequence as described above; for each pixel, there is therefore a corresponding decay curve.  However, before processing these curves as described in detail above, we first apply the NESMA filter to the stack of images, resulting in pixel-wise decay curves of greatly improved SNR. These are then inverted using SpanReg to obtain a $T_2$ distribution for each pixel. The reference MWF map is formed from these distribution as described above, by taking the integral of the DF's up to $T_2$=40 ms for each pixel separately. 

The MWF maps created from high and low SNR data for comparison, shown in the same figure, were generated as follows. The high-quality $T_2$ distribution functions used to define the reference map were used to generate noiseless decay curves, to which Gaussian noise was added to achieve SNR$=800$ (high SNR) and SNR$=200$ (low SNR) in the decays. The decay curve at each pixel to corresponds to $\yob$ in Eq. \eqref{fwd_eqn}. We then reconstructed the $T_2$ DF's using SpanReg, and NNLS with DP regularization and without regularization. MWF maps were constructed from these and compared to the reference map using the SAD metric, Eq. \eqref{scaled_absolute_difference}.  

Visual inspection shows that the SpanReg analysis yields a result more closely resembling the reference image than either of the other two methods. 
Quantitatively, the scaled absolute difference (SAD) between the reference map and the derived MWF maps from the high SNR dataset are: 0.32 (compared to SpanReg), 0.38 (DP), and 0.59 (NNLS). The corresponding values for the low SNR dataset are: 0.39 (SpanReg), 0.58 (NNLS with DP regularization), and 0.68 (non-regularized NNLS). Thus, by this metric, SpanReg outperforms DP for the high SNR and low SNR datasets by $16\%$ resp. $33\%$, and NNLS by $46\%$ resp. $43\%$. 

\subsubsection{Comparison of Performance for High Versus Low SNR Data}
\label{experiments_results}
We now examine the sensitivity to noise of the reconstruction methods. Even with spatially invariant noise, SNR varies across an image because of signal strength variations. Ideally the entire SpanReg algorithm would be applied separately for each pixel according to its SNR. This is not practical for typical images; in our case, the images consist of $\sim 25,000$ pixels.  Therefore, we divided the full range of image SNR into 18 bins over the range of 10 to 800, encompassing both unfiltered and NESMA-filtered pixel-wise decay signals. The same Gaussian basis functions were used in all cases, but different SpanReg parameter values and functions $\lc\beta_{i,\lambda_j}\rc$, $\lc\bg_{i,\lambda_j}\rc$ were obtained for each of these 18 SNR bins based on its SNR.  The corresponding $\lc\beta_{i,\lambda_j}\rc$ and $\lc\bg_{i,\lambda_j}\rc$ were used to reconstruct the data from each pixel with SNR within that bin. MWF for each pixel was then derived from the resulting distributions as described above, by integrating the DF's up to 40 ms. The resulting MWF maps are displayed in the Supplementary Information, Fig. \ref{MWF_invivo}.

We first note that the results for SpanReg are visually more similar at the two noise levels than are the results for the other reconstruction methods, indicating greater stability with respect to noise. In addition, the MWF distributions recovered by DP and LS show prominent edge effects surrounding the ventricles, as well as myelin voids towards the brain periphery. In contrast, the MWF maps obtained with SpanReg show a much more qualitatively normative pattern. Quantitatively, the SAD between the high and low SNR MWF maps, $SAD\lp \bM_{\text{high SNR}} , \bM_{\text{low SNR}} \rp$, for the three reconstruction methods are 0.35 for SpanReg, 0.43 for DP, and 0.63 for NNLS. This is consistent with the results shown in Section \ref{simulation_brain}.

\section{Discussion}
In this paper, we present a new approach, SpanReg, to obtaining a regularized solution of the discretized Fredholm equation of the first kind. This technique should be widely applicable for inverse problems requiring regularization to obtain meaninful solutions. The idea of the method is to incorporate information content over a range of regularized solutions rather than to attempt to identify a single optimal regularization parameter, $\lambda$.  For inverse problems of a type arising extensively in MR relaxometry, we provided simulations demonstrating greater accuracy of SpanReg as compared to conventional Tikhonov regularization using the DP for selection of $\lambda$. We also showed that SpanReg is more stable with respect to the choice of regularization parameters, in the sense that the departure from an optimal solution is smaller when the solution arrived at using the optimal $\lambda$ is not incorporated. The method was also demonstrated on experimental biomedical MRR data to evaluate the MWF in the human brain; this is a very challenging problem of great current interest.

The framework put forward in this work is to provide a more general form of the recovered DF: $\f^* = \sum_j \alpha_j \f_{\lambda_j}$; conventional Tikhonov regularization incorporates just a single term, optimal in some sense, of this sum. In addition to relaxing the restriction of isolating the single best regularized solution, our method also provides a flexible form for the construction of the desired solution. To obtain the coefficients $\lc\alpha_j\rc$, we formulated the solution $\sum_j \alpha_j \f_{\lambda_j}$ as a linear combination of multiple regularized solutions via an over-complete dictionary of Gaussian functions $\lc \bg_{i,\lambda_j} \rc$. The algorithm can be separated into offline and online computations, where the offline part is specific to a given SNR and basis set, but is independent of a specific instance of experimental data. The offline calculation requires the determination of the noise-corrupted basis functions and parameters, e.g. fixing the number $M$ of Gaussian dictionary functions $\bg_{i}$, the number $N$ of regularization parameters used by the algorithm, and performing the actual computation required to obtain the noise-corrupted DF, $\lc\bg_{i,\lambda_j}\rc$.The online calculation is dependent upon a specific data set and is much less computation-intensive. 

This formulation exhibits a condition number $cond(\bB) = O(10^5)$ in Eq. \eqref{online_explicit_form_ls} for the simulation analysis in Section \ref{sec: SpanReg_vs_dp}, compared to the condition number of $cond(\bA)$ in the NNLS problem, which is $O(10^{20})$. 

In this study, we demonstrated that the SpanReg method achieves higher accuracy than the conventional DP parameter selection method, which also requires knowledge of noise level in the observation $\yob$. For the prototypical bimodal DF's studied, SpanReg exhibits a substantially greater ability to resolve the two components. With more disparate spaced components, SpanReg also achieved improved reconstruction of positions and amplitudes of the components. Further, SpanReg is not constrained to the selection of a single optimal value of $\lambda$; in fact, in conventional practice, this optimal value will depend upon the specific selection criterion implemented. Thus, application of different conventional methods is virtually guaranteed to recover different DF's. In contrast, our more general formulation forms the reconstruction through a linear combination of differently-weighted solutions with appropriate weights, and so is less dependent upon the exact degree of regularization of any one of these. 

Previous studies have also presented important reconstruction methods based on Tikhonov regularization with multiple regularization parameters. In distinct contrast to our work,these have employing a modified regularization penalty from which is derived a single optimal solution.  These methods include multiple regularization \cite{Belge_2002, Gopi_2014}, multi-parameter regularization \cite{Chen_2008, lu_2011, Wang_2012, gazzola_2016,  Chung_2017}, and multi-penalty regularization \cite{Fornasier_2014, ABHISHAKE_2016}. This literature generally studies the same optimization problem as we have worked with, with the somewhat more general form:
\begin{equation}
    \label{multiple_reg}
    \f^* = \argmin_{\f}\lc \norm{\yob - \bA\f}_2^2 + \sum_{j = 1}^N \lambda_j \mathcal{L}_j\lp\f\rp\rc
\end{equation}
where $\mathcal{L}_j$ are regularization operators and $\lambda_j$ are regularization parameters associated with the corresponding operator. One of the most well-known of these is elastic-net regularization (EN) \cite{Zou_2005}, where $\mathcal{L}_1\lp\f\rp = \norm{\f}_1$, and $\mathcal{L}_2\lp\f\rp = \norm{\f}_2^2$. The EN is able to promote sparsity while also being able to handle smoother distributions. Other types of penalty terms have also been proposed involving smoothness constraints; these include penalties of the form $\norm{\nabla \f}_1$ \cite{Belge_2002, lu_2011, Wang_2012, Gopi_2014, gazzola_2016, Chung_2017}, $\norm{\nabla\f}_2^2$ \cite{lu_2011, Fornasier_2014, gazzola_2016}, and other penalties of the form $\norm{\bW\f}$ \cite{Belge_2002, Chen_2008, Gopi_2014, Chung_2017}, where $\bW$ may be, for example a projection or transformation matrix. Unlike SpanReg, all of these methods seek optimal regularization parameters for a specific form of regularization rather than incorporating differently-regularized solutions. 

An interesting alternative method more closely related to ours deviates from the formalism of Eq. \eqref{multiple_reg} by constructing a linear combination of regularized solutions \cite{HOCHSTENBACH_2012}, as in our treatment. The components forming the recovered DF are obtained from different regularization methods, with, as in conventional treatments and distinct from SpanReg, a single optimal regularizer being identified for each. Nevertheless, this work introduces the idea of combining DF's, each of which in effect reflects a potentially desirable quality, into a final derived DF.

A variety of parameter selection techniques have been presented for the conventional types of multi-parameter regularization described above, generally based on the classical approaches for single-parameter regularization optimization. These include the multi-parameter discrepancy principle \cite{lu_2011}, L-hypersurface \cite{Belge_1998, Belge_2002}, and GCV-multi \cite{Chung_2017}. Other methods \cite{Chen_2008} such as simple grid search \cite{sabett2017l1} and the balancing principle \cite{ABHISHAKE_2016} have been implemented. However, as emphasized above, none of these studies presents the notion of including sub-optimal regularized solutions as we have proposed and demonstrated with SpanReg. Thus, the current work is fundamentally different from what has appeared in the literature in; we form a linear combination of sub-optimal solutions, in contrast to previous work which has used combinations of regularization terms leading to a single ``optimal" solution, or \cite{HOCHSTENBACH_2012} a linear combination of optimal solutions. SpanReg allows us to combine the stability of more highly-regularized solutions with the resolution of less-regularized solutions; the potential for the efficacy of this construction was provided by our observations of the different responses to regularization of very similar ill-posed problems.
 
SpanReg should be widely applicable to a wide range of linear inverse problems based on the Fredholm equation of the first kind. As one class of examples, we demonstrated its applicability to MRR investigations of brain, where the DF is based on the $T_2$ distribution of tissue. However, the same formulation also applies to diffusion MRI \cite{Benjamini_Komlosh_Basser_2017}, where application is simpler due to the availability of the $b=0$ data point, where $b$ is the conventional symbol for the combination of gradient strengths and timing that defines diffusion sensitization. In this case, the formalism of Eq. \eqref{TE=0_formula} is not required. Similar comments apply to the closely-related problem of $T_1$ MRR \cite{HURLIMANN2006303}. Further, MRR applications extend far beyond those in biomedicine, and include the food sciences \cite{kirtil20161}, engineering \cite{Greener_Peemoeller_Choi_Holly_Reardon_Hansson_Pintar_2000}, and the petrochemical industry \cite{SONG201917}. There are also many applications of SpanReg outside of MR studies, including fluorescence analysis \cite{Reisser_Hettich_Kuhn_Popp_Grosse-Berkenbusch_Gebhardt_2020}.  

Although we have presented a novel and effective method for combining multiple degrees of regularization into an ill-conditioned problem, certain limitations remain.  We have not fully explored selection of an optimal basis, although the Gaussian basis set used has been shown to perform well for reconstructing a range of challenging DF's. 
In addition, there are a number of user-selected parameters for SpanReg, and we have not yet explored these systematically. These include the number $N$ of regularized solutions to incorporate into the DF as well as their associated regularization parameter values $\lambda_j$; the size $M$ and details of the Gaussian dictionary used to represent the DF for each of these solutions; and the number $n_{\text{run}}$ of noise realizations used for averaging to render the results robust with respect to noise realization. In this paper, our considerations have been based on the goal of maintaining reasonable conditioning of the the LS problem in Eq. \eqref{eqn_for_beta}; as demonstrated, this empirical approach has worked well. 
We also emphasize that the requirement for knowledge of SNR is a limitation of SpanReg in many contexts, although for the large category of problems based on decaying signals, including MRR and related experiments in MR, SNR estimates are generally available. This consideration also applies to certain conventional parameter selection methods such as DP. 
In spite of these limitation, we have shown that SpanReg can be applied to datasets with varying, though known, levels of SNR, such as in the brain MWF analysis. This does require a lengthier offline computation, with separate calculations required across a range of SNR values. The degree to which a given range of SNR values must be discretized also remains an open question, though the success of the binning procedure described in the brain MRI analysis support the notion that this discretization need not be unduly fine. This is a critical finding, indicating that the noise estimate required for SpanReg can be approximate. The degree to which this holds remains to be quantified.

In addition to theoretical developments and demonstration on simulated data, we have provided an analysis of in vivo brain MRI data. We compare the performance of SpanReg to two other methods for recovering DF's from decay curves for application to brain MWF mapping. This presentation is not intended to provide an extensive comparison of SpanReg with other state-of-the-art methods for assessing MWF; development of these methods is a topic of major current interest, with rapid introduction of new approaches \cite{mackay2016magnetic, BOUHRARA2017800, Bouhrara_NESMA_2018}. A full comparison with these is beyond the scope of the present manuscript, which is to introduce and provide an initial application of SpanReg.

In addition to addressing the current limitations of SpanReg as outlined above, potential extensions include exploration of alternative forms of Tikhonov and other forms of regularization penalties. 
In addition, SpanReg can be extended in a straightforward fashion to 2D and higher dimensional MRR analysis \cite{MITCHELL201234}, for which the governing inverse problem is a higher-dimensional inverse Laplace transform. 

In conclusion, we have proposed a new approach to determining regularized solutions to the Fredholm integral equation of the first kind by incorporating the information content of non-optimal solutions, and have demonstrated its efficacy in simulations and through application to MRR of the human brain. SpanReg should be widely applicable throughout the field of inverse problems, presenting an alternative to Tikhonov and related forms of regularization
\section*{Acknowledgments}
This work was supported by the Intramural Research Program of the National Institute on Aging of the National Institutes of Health. We thank Pak-Wing Fok for particularly helpful discussions.

\bibliographystyle{plain}

\section{Supplementary information}

\begin{figure}[H]
\centering
\captionsetup{width=.8\linewidth}
\includegraphics[width=0.9 \linewidth]{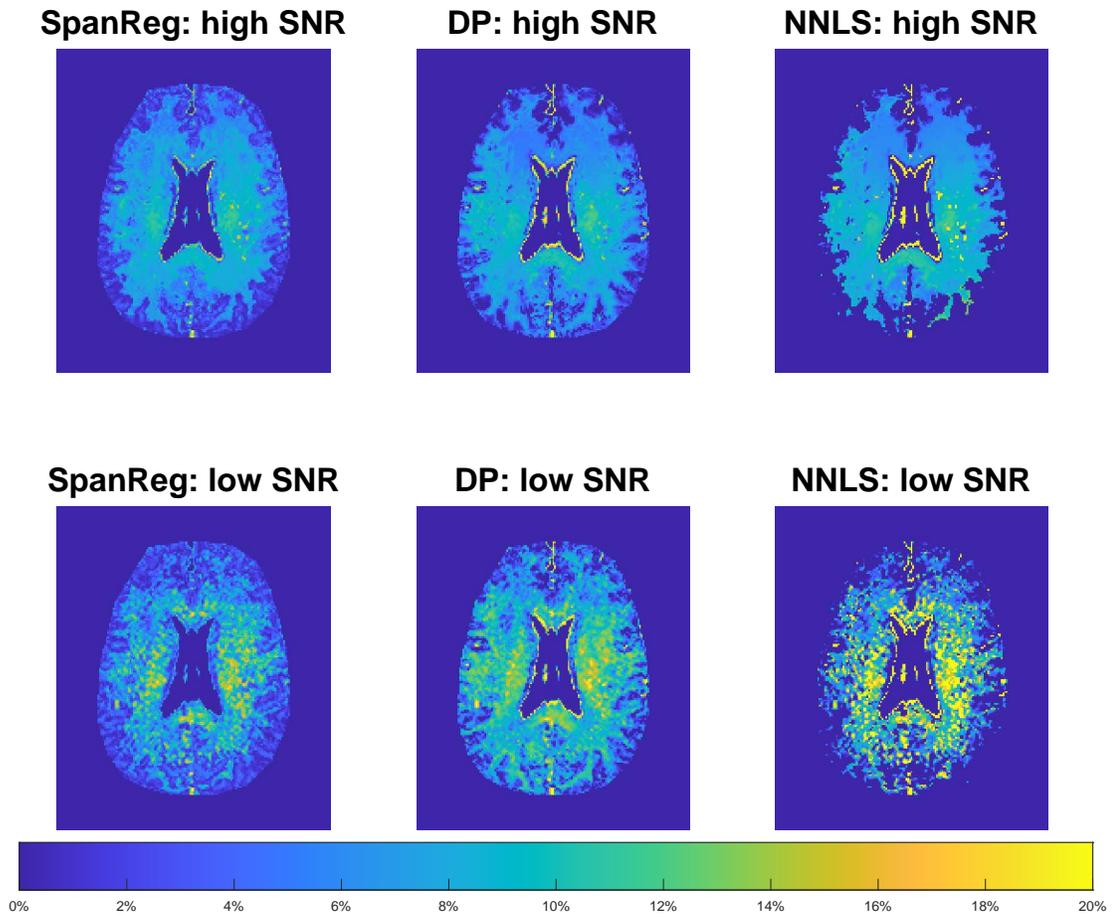}
\caption{MWF maps reconstructed from original imaging data, to which two different levels of noise were added.}
\label{MWF_invivo}
\end{figure}

\end{document}